\documentclass[aps,prb,twocolumn,superscriptaddress,amsfonts,amsmath,amssymb]{revtex4}
\usepackage{bm}
\usepackage{graphicx}
\usepackage{amsmath}
\usepackage{amstext}
\usepackage{amssymb}
\usepackage{amsfonts}
\usepackage{amsbsy}
\usepackage{verbatim}

\def\ba{\boldsymbol{a}}
\def\bz{\boldsymbol{z}}
\def\br{\boldsymbol{r}}
\def\bx{\boldsymbol{x}}
\def\bp{\boldsymbol{p}}
\def\bR{\boldsymbol{R}}
\def\bk{\boldsymbol{k}}
\def\bK{\boldsymbol{K}}
\def\bQ{\boldsymbol{Q}}
\def\bq{\boldsymbol{q}}
\def\bS{\boldsymbol{S}}
\def\bT{\boldsymbol{T}}
\def\rU{{\rm U}}
\def\rU1{{\rm U}(1)}
\def\rSU2{{\rm SU}(2)}
\def\er{\epsilon_{\br}}
\def\bJ{\boldsymbol{J}}

\def\bL{\boldsymbol{L}}
\def\balpha{\boldsymbol{\alpha}}
\def\bsigma{\boldsymbol{\sigma}}
\def\btau{\boldsymbol{\tau}}
\def\blambda{\boldsymbol{\lambda}}
\def\tr{\operatorname{tr}}
\def\bpperp{\boldsymbol{p}_{\perp}}
\def\Dslash{\not \!\! D}

\begin{document}

\title{$\rSU2$ gauge theory of the Hubbard model and application to the honeycomb lattice}
\author{Michael Hermele}
\affiliation{Department of Physics, Massachusetts Institute of Technology, Cambridge, Massachusetts 02139, USA}

\date{\today}
\begin{abstract}
Motivated by recent experiments on the triangular lattice Mott-Hubbard system $\kappa$-(BEDT-TTF)$_2$Cu$_2$(CN)$_3$, we develop a general formalism to investigate quantum spin liquid insulators adjacent to the Mott transition in Hubbard models.  This formalism, dubbed the $\rSU2$ slave-rotor formulation, is an extension of the $\rSU2$ gauge theory of the Heisenberg model to the case of the Hubbard model.  Furthermore, we propose the honeycomb lattice Hubbard model (at half-filling) as a candidate for a spin liquid ground state near the Mott transition; this is an appealing possibility, as this model can be studied via quantum Monte Carlo simulation without a sign problem.  The pseudospin symmetry of Hubbard models on bipartite lattices turns out to play a crucial role in our analysis, and we develop our formalism primarily for the case of a bipartite lattice.  We also sketch its development for a general Hubbard model.  We develop a mean-field theory to describe spin liquids and some competing states, and apply it to the honeycomb lattice.  On the insulating side of the Mott transition, we find an $\rSU2$ algebraic spin liquid (ASL), described by gapless $S = 1/2$ Dirac fermions (spinons) coupled to a fluctuating $\rSU2$ gauge field.  This result contrasts with that obtained via a $\rU1$ slave-rotor approach, which instead found a $\rU1$ ASL.  That formulation does not respect the pseudospin symmetry, and is therefore not correct on the honeycomb lattice.  We construct a low-energy effective theory describing the ASL phase, the conducting semimetal phase and the Mott transition between them.  This physics can be detected in numerical simulations via the simultaneous presence of substantial antiferromagnetic and valence-bond solid correlations.  In the $\rSU2$ ASL, these observables have slowly-decaying fluctuations in space and time, described by power laws with the same critical exponent.  We recover the results of the $\rU1$ slave-rotor formulation in the presence of a strong breaking of pseudospin symmetry.  Our analysis suggests that both a third-neighbor electron hopping, and/or pseudospin-breaking terms such as a nearest-neighbor density interaction, may help to stabilize a spin liquid phase.
\end{abstract}
\maketitle

\section{Introduction}
\label{sec:intro}

In the quest for a quantum spin liquid Mott insulator in two or more spatial dimensions,\cite{anderson87} most of the theoretical attention to date has focused on quantum magnets.  While this effort has resulted in significant progress in our general understanding of a variety of different spin liquids, as well as some simple models that have been shown to realize such ground states (see Refs.~\onlinecite{sachdev04,senthil04,palee06, alet06}  for reviews), 
spin liquids may be rather rare in the simplest quantum magnets.  In order to advance the understanding in this area, more models and candidate materials are urgently needed, and it is important to look to other classes of systems.

Following a series of recent experiments, the organic material 
 $\kappa$-(BEDT-TTF)$_2$Cu$_2$(CN)$_3$ has emerged as the first clear candidate for a spin liquid ground state.\cite{shimizu03, kawamoto04, kurosaki05}  It is believed that this material can be described in terms of a single-band triangular lattice Hubbard model at half-filling (one electron per site).  The experimentally observed spin liquid lies just on the insulating side of a metal-insulator transition; this transition and the metallic state across it, which becomes a superconductor at low temperatures, can be accessed by applying pressure.  The interpretation in terms of the Hubbard model is that applying pressure tunes $t/U$ (the ratio of kinetic energy to Coulomb repulsion) and drives the system across the Mott transition.  Among the many striking features observed thus far, it is notable that the spin liquid physics appears adjacent to the Mott transition; this is natural, as local charge fluctuations tend to disrupt magnetic ordering, which in turn is more likely to be favored for stronger correlation (smaller $t/U$).  The message is that, when the quantum nature of $S = 1/2$ spins and geometrical frustration are not enough to disrupt magnetic ordering, as is likely the case for the triangular lattice \emph{Heisenberg} model,\cite{capriotti99} local charge fluctuations may be able to tip the balance toward a spin liquid state.  We are therefore optimistic that more examples of spin liquid physics are waiting to be found near the Mott transition in other systems.

The primary subject of this paper is the development of a general formalism to describe spin liquids adjacent to the Mott transition in Hubbard models.
At a minimum, we wish to describe the simplest scenario, namely a direct (and continuous) quantum phase transition between a spin liquid Mott insulator and a conventional conducting state, which may be a metal or a superconductor.   
What features should the necessary formalism possess?  First, a description starting from the level of mean-field theory should be possible.  Generally, spin liquids can be described in terms of $S = 1/2$ spinons, which may be either bosons or fermions, carrying zero electromagnetic charge and coupled to an emergent gauge field.  In the present case a description in terms of fermionic spinons is preferable, as this allows a mean-field description of the Mott transition.  
Therefore, a natural starting point is to split the electron into a fermionic spinon and a bosonic degree of freedom carrying the charge.  

For our formalism to be as general as possible, it should tie in to the fermionic-spinon theory of the \emph{Heisenberg} model,\cite{wen02} which is appropriate for  describing a large class of spin liquids in the large-$U/t$ limit.  In particular, any spin liquid state that can be described in the Heisenberg model should also be present in our Hubbard model formalism.  
In all slave-particle formulations a local gauge redundancy is introduced; in the fermionic-spinon theory of quantum magnets this is an $\rSU2$ redundancy.\cite{affleck88a, dagotto88}
In order to satisfy the above criterion, our formulation of the Hubbard model should also possess an $\rSU2$ gauge redundancy.  Therefore, we need to generalize the $\rSU2$ gauge theory of the Heisenberg model to the Hubbard model.  
This program is in the same spirit as the $\rSU2$ gauge theory of the $t$-$J$ model, which allows for a description of \emph{doped} spin liquids.\cite{wen96,palee98}  The path pursued here allows instead for a description of spin liquids with local charge fluctuations, as appropriate for insulating phases of the half-filled Hubbard model.

A formalism \emph{partially} satisfying the above criteria has already been achieved in the $\rU1$ slave-rotor formulation.\cite{florens04,sslee05}  In this approach the electron is formally split into a product of a fermionic spinon and a charge-carrying rotor boson.  There is only a local $\rU1$ redundancy, however, and only a subset of possible spin liquid phases can be described (at the mean-field level) without extending the formalism; in particular, states with spinon pairing and a $Z_2$ emergent gauge field, or those where the $\rSU2$ gauge symmetry is unbroken, are not accessible.

Here, we build on the $\rU1$ slave-rotor formulation and develop the more general $\rSU2$ slave-rotor formulation, which satisfies the above criteria for a general description of spin liquids near the Mott transition.  Introducing an $\rSU2$ gauge redundancy, the electron is split into a product of a fermionic spinon and an ``$\rSU2$ matrix rotor.''  While this formalism should be useful quite generally, and we do sketch its development for a general Hubbard model here, we develop it primarily in the context of the honeycomb lattice Hubbard model (at half-filling), where it turns out to be not only useful but in fact \emph{essential}.  This is due to the pseudospin symmetry of Hubbard models on bipartite lattices;\cite{cnyang89,zhang90}
as discussed in more detail below, this symmetry is explicitly (and inappropriately) broken by the $\rU1$ slave-rotor formulation, but is respected by the $\rSU2$ formulation.  This turns out to have important consequences for the physics near the Mott transition.

The motivation to consider the honeycomb Hubbard model comes largely from the present understanding of the  $\kappa$-(BEDT-TTF)$_2$Cu$_2$(CN)$_3$ system.  Recent theoretical work has proposed that the spin liquid state can be described in terms of a Fermi surface of neutral spin-1/2 spinons coupled to an emergent $\rU1$ gauge field.\cite{motrunich05, sslee05, motrunich06, sslee06}  Also, microscopic studies of the triangular Hubbard model find a spin-liquid insulator adjacent to the Mott transition,\cite{kashima01,morita02} although the accuracy of the approximations involved is not clear.
These ideas are quite promising as an explanation of the experimental data.  However, two obstacles to further progress exist in this system, and our understanding would be greatly advanced if a complementary system exhibiting similar physics but not suffering from these problems can be found.
  First, the spinon Fermi surface state is relatively complicated, and our understanding of it is probably no so well-developed compared to that of other spin liquid states.  Second, the triangular lattice Hubbard model cannot be effectively studied using quantum Monte Carlo simulation due to the notorious sign problem.

The honeycomb lattice Hubbard model is then an excellent candidate to complement the understanding of the triangular lattice system.  First, because the honeycomb lattice is bipartite, the model can be studied by quantum Monte Carlo simulation without a sign problem.  Also, the lack of nesting implies that a metal-insulator transition must be present at nonzero $U/t$, unlike in the square lattice.  Second, it turns out that the most likely candidate for a spin liquid is an example of an \emph{algebraic spin liquid}
(ASL),\cite{affleck88,marston89,rantner01,rantner02}  which has only spinon Fermi points rather than a full Fermi surface.  Such spin liquids are a good deal simpler than the spinon Fermi surface state, and several recent works have significantly solidified the theoretical
understanding.\cite{rantner01,rantner02,borokhov02,hermele04,hermele05}  Furthermore, it is natural to ask whether a spin liquid is possible in the honeycomb model, as it lacks magnetic frustration.  We speculate that magnetic frustration is not a necessary condition for a spin liquid ground state, and that the nontrivial competition between kinetic and potential energy near the metal-insulator transition can be enough on its own.  The honeycomb model provides a simple system in which to address this issue.  At present, we do not know of a strongly correlated material described by the honeycomb Hubbard model; the results of this paper make it clear that such materials could be very interesting, if they can be fabricated.

\begin{figure}
\includegraphics[width=3in]{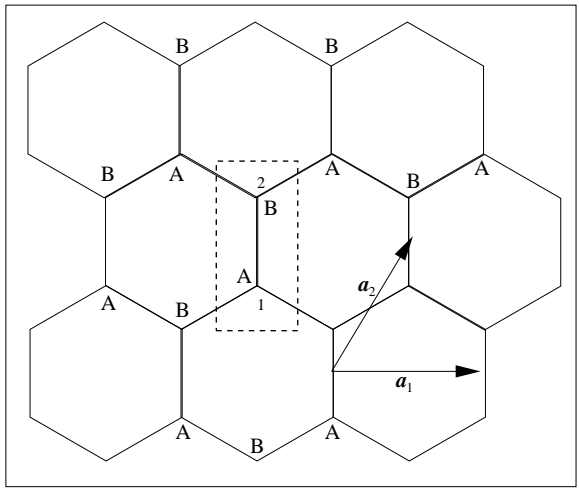}
\caption{The honeycomb lattice with $A$ and $B$ sublattices labeled.  The unit cell consists of the two sites (labeled 1 and 2) inside the dashed-line box.  Unit cell positions are specified by Bravais lattice 
vectors $\bR = n_1 \ba_1 + n_2 \ba_2$, where $n_1, n_2$ are integers. The primitive vectors $\ba_1 = (1,0)$ and $\ba_2 = (1/2, \sqrt{3}/2)$ are shown in the figure.}
\label{fig:honeycomb}
\end{figure}

The Hamiltonian of the honeycomb lattice Hubbard model is
\begin{eqnarray}
\label{eqn:hubbard-model}
{\cal H}_{{\rm H}} &=& - \sum_{( \br, \br' )} \big( t_{\br \br'} c^\dagger_{\br \alpha} c^{\vphantom\dagger}_{\br' \alpha} + \text{H.c.} \big) \\
&+& U \sum_{\br} (c^\dagger_{\br \uparrow} c_{\br \uparrow} - \frac{1}{2} )
(c^\dagger_{\br \downarrow} c_{\br \downarrow} - \frac{1}{2} ) \text{.} \nonumber
\end{eqnarray}
Here, $\br$ labels the sites of the honeycomb lattice (shown in Fig.~\ref{fig:honeycomb}), $c_{\br \alpha}$ destroys an electron on site $\br$ with spin $\alpha$, the sum in the first term is over all pairs of sites $(\br, \br')$, and repeated spin labels are summed over.  We sometimes find it convenient to label honeycomb lattice sites by the pair $(\bR, i)$, where $\bR$ is a Bravais lattice vector, and $i = 1,2$ labels the two-site basis of the unit cell as shown in Fig.~\ref{fig:honeycomb}.  The honeycomb lattice is bipartite, and in order to preserve this structure we require $t_{\br \br'} = 0$ when $\br$ and $\br'$ lie in the same sublattice.  In addition to the nearest-neighbor hopping $t$, we will also discuss the possible effect of the third-neighbor hopping $t''$. We restrict our attention entirely to the case of a half-filled band; combined with the bipartite structure this ensures particle-hole symmetry, which is required for sign-problem-free Monte Carlo simulation.  Our goal here is to understand the simplest scenarios for spin liquid physics near the Mott transition, and to make predictions that can be tested in future numerical simulations.  One quantum Monte Carlo study of the honeycomb Hubbard model has already been carried out;\cite{paiva05} see Sec.~\ref{sec:discussion} for a discussion.

This model has a pseudospin $\rSU2$ symmetry, which will play a crucial role in our analysis and is reviewed in more detail in Sec.~\ref{sec:pseudospin}.  On a single site of the lattice, pseudospin rotates between the empty and doubly-occupied states, leaving the $S = 1/2$ singly-occupied states unaffected.  
It will be useful to contemplate adding the nearest-neighbor density interaction
\begin{equation}
\label{eqn:hv-term}
{\cal H}_V = V \sum_{\langle \br \br' \rangle} (c^\dagger_{\br \alpha} c^{\vphantom\dagger}_{\br \alpha} - 1) (c^{\dagger}_{\br' \alpha} c^{\vphantom\dagger}_{\br' \alpha} - 1) \text{.}
\end{equation}
This term breaks pseudospin down to an ${\rm O}(2)$ subgroup generated by $\rU1$ charge rotations and particle-hole symmetry.  

The model Eq.~(\ref{eqn:hubbard-model}) has a stable metallic phase for small-$U/t$, and the large-$U/t$ ground state is expected to be an insulating N\'{e}el antiferromagnet (AF).  Actually, the metallic state is a semimetal (SM) with Dirac nodes instead of a full Fermi surface, and a linearly-vanishing density of states at the Fermi energy.  At intermediate-$U/t$ there must therefore be a metal-insulator transition, with the possibility of spin liquids or other interesting physics nearby.  This situation contrasts dramatically with the square lattice model, which, at half-filling and in the absence of intra-sublattice hopping, is believed to be in the AF phase for all values of $U/t$, due to the nested Fermi surface.

\begin{figure}
\includegraphics[width=3in]{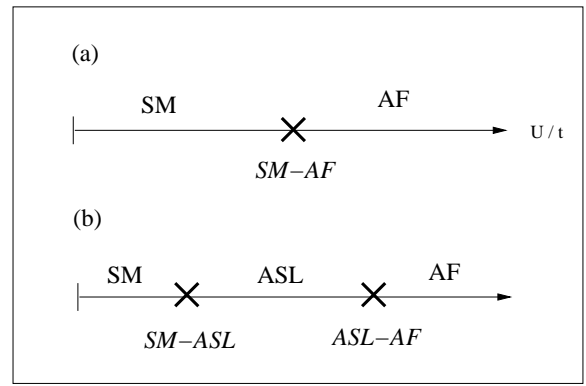}
\caption{Two scenarios for the zero-temperature phase diagram of the honeycomb Hubbard model as a function of $U/t$.  In (a), there is a direct transition between the semimetal (SM) and N\'{e}el antiferromagnet (AF).  This SM-AF transition does not involve spin liquid physics.  Another possibility, shown in (b), is that a region of algebraic spin liquid (ASL) intervenes between SM and AF.}
\label{fig:scenarios}
\end{figure}

We caution that the very simplest possibility for the metal-insulator transition actually has nothing to do with spin liquid physics, and can be understood by coupling a fluctuating N\'{e}el order-parameter field to the gapless Dirac fermions of the semimetal.  This results in a direct, continuous transition between SM and AF, which can be analyzed in a suitable large-$N$ or $4 - \epsilon$ expansion.\cite{herbut06}  The phase diagram in this scenario is shown in Fig.~\ref{fig:scenarios}a.  While this critical point is already interesting, as the coupling to the gapless fermions changes the critical exponents from their ${\rm O}(3)$ values,\cite{herbut06} we focus on the even more exciting possibility of spin liquid physics near the Mott transition.

The main result of our analysis is the possibility of a Mott transition from the semimetal to an \emph{algebraic spin liquid} Mott insulator.\cite{affleck88,marston89,rantner01,rantner02}  The resulting phase diagram is shown in Fig.~\ref{fig:scenarios}b.  Two transitions are required as $U/t$ is increased:  first, the Mott transition from the semimetal to the ASL, and then a transition from the ASL to the N\'{e}el antiferromagnet; an example of the latter transition has been studied in Ref.~\onlinecite{ghaemi06}.
The ASL, which is a critical state with power-law correlations for many observables and no spin gap, can be thought of as a two-dimensional analog of the ground state of the $S = 1/2$ Heisenberg chain.  It is an interacting critical state in the sense that the fermionic spinons are not free quasiparticles even at the lowest energies.  It can be described by a field theory of massless Dirac fermions coupled to a gauge field, and can roughly be thought of as the deconfined phase of the gauge theory.  Actually there are two different kinds of ASL, depending on whether the gauge field is $\rU1$ or $\rSU2$ -- in the present case we find an $\rSU2$ ASL.  Both kinds of ASL can be analyzed reliably in the limit of a large number of Dirac fermion flavors (large-$N_f$ limit),\cite{appelquist86,appelquist90,rantner01,rantner02, vafek02, franz02, hermele04, hermele05,ran06}  and have nontrivial critical exponents that vary as a function of $N_f$.  The $\rU1$ ASL is a stable phase in the large-$N_f$ limit;\cite{rantner01,rantner02,hermele04,borokhov02} this is expected also to be true of its $\rSU2$ counterpart.\cite{ran06}

For simplicity, in this paper we shall assume that the $\rSU2$ ASL is a stable phase.  While this is true in the large-$N_f$ limit, it need not be the case for the Hubbard model.  Our results may be relevant to the Hubbard model even if the $\rSU2$ ASL is unstable; this is briefly discussed in Sec.~\ref{sec:discussion}.

Our results clarify a brief discussion in Ref.~\onlinecite{sslee05}, which focused on the Mott transition in the triangular lattice Hubbard model, but also considered the honeycomb lattice model using the $\rU1$ slave-rotor formulation. That analysis missed the role of pseudospin symmetry and found a Mott transition to the $\rU1$ ASL.  Here, the $\rSU2$ slave-rotor formulation has allowed us to properly take the pseudospin symmetry into account, and instead we find a Mott transition to a $\rSU2$ ASL.  This is not just a subtle distinction; the $\rU1$ and $\rSU2$ ASL have distinct physical properties, and it should be rather easy to distinguish them in a numerical simulation.
If, however, the pseudospin symmetry is broken
 -- by adding ${\cal H}_V$, for example -- a Mott transition to the $\rU1$ ASL again becomes possible.
 Based on the intuition that gauge theories with more colors are more susceptible to confinement, the $\rU1$ ASL is likely to be more stable than its $\rSU2$ counterpart.  Therefore, in order to observe spin liquid physics in numerical simulations it may actually be helpful to include ${\cal H}_V$.
We stress that the issue of pseudospin does not affect the results of Ref.~\onlinecite{sslee05} for the triangular lattice model, where pseudospin symmetry is absent.  However, it will still be interesting to apply the $\rSU2$ formulation there, as it addresses the competition among a wider range of spin liquids.

Another of our major results is a suggestion for how to \emph{detect} the $\rSU2$ ASL in numerical simulations.  Very recently, Ran and Wen have shown that the $\rSU2$ ASL has an emergent ${\rm Sp}(4)$ symmetry at low energies, which contains spin-rotation symmetry as a subgroup.\cite{ran06}  This work was done in the context of the $\pi$-flux spin liquid on the square lattice, which has the same low-energy theory as the $\rSU2$ ASL studied here.  Furthermore, Ref.~\onlinecite{ran06} showed that the ${\rm Sp}(4)$ symmetry unifies the N\'{e}el vector and the two-component order parameter for columnar and box valence-bond solid (VBS) states\cite{read90} into a five-component vector.  These observables have slowly-decaying power law correlations, all with the same critical exponent.  These results are analogous to those of Ref.~\onlinecite{hermele05} on ${\rm SU}(4)$ emergent symmetry and a unification of various ``competing orders'' in the $\rU1$ ASL.  In the honeycomb case we also find a unification of the N\'{e}el vector with the VBS order shown in Fig.~\ref{fig:hcvbs}.  Therefore, we propose to look for the $\rSU2$ ASL in numerical simulations by testing for the simultaneous presence of slowly decaying N\'{e}el and VBS correlations (see Sec.~\ref{sec:spin-liquid-ft} for further discussion).

We now give an outline of the paper and sketch our approach in some more detail.  First, in Sec.~\ref{sec:earlier-work}, we briefly discuss connections to some earlier work.
Following a brief discussion of pseudospin symmetry (Sec.~\ref{sec:pseudospin}),  we review the $\rU1$ slave-rotor formulation in Sec.~\ref{sec:u1-attempt}; there, the electron is split into a fermionic spinon and a charge-carrying ${\rm O}(2)$ rotor.  We show that the $\rU1$ formulation explicitly breaks the pseudospin symmetry and is thus inadequate to describe the honeycomb Hubbard model.  We are thus led to introduce the $\rSU2$ slave-rotor formulation in Sec.~\ref{sec:su2-formulation}.  
This is done by splitting the electron into fermionic spinons and an $SU \textit{(2)}$ \emph{matrix rotor}.  In contrast to the ${\rm O}(2)$ rotor, which can be thought of as a quantum particle constrained to move on a ring, this rotor is a particle constrained to move in the space of $\rSU2$ matrices (or, equivalently, on the 3-sphere).  Our formulation is obtained by first organizing the electron operators into a $2 \times 2$ matrix
\begin{equation}
\label{eqn:Psi-defn}
\Psi_{\br} = \left( \begin{array}{cc}
c_{\br \uparrow} & \epsilon_{\br} c^\dagger_{\br \downarrow} \\
c_{\br \downarrow} & -\epsilon_{\br} c^\dagger_{\br \uparrow} \end{array} \right)
\text{,}
\end{equation}
where $\epsilon_{\br} = \pm 1$ for $\br$ lying in the $A$/$B$ sublattice (see Fig.~\ref{fig:honeycomb}).  This matrix is designed so that left-multiplication by an $\rSU2$ matrix is a spin rotation, while pseudospin rotation is achieved by \emph{right}-multiplication by an $\rSU2$ matrix.  We therefore split
$\Psi_{\br}$ into the product
\begin{equation}
\label{eqn:splitting-Psi}
\Psi_{\br} = F_{\br} Z_{\br} \text{,}
\end{equation}
where $F_{\br}$ is a $2 \times 2$ matrix of fermionic spinon operators $f_{\br \alpha}$ defined just as in Eq.~(\ref{eqn:Psi-defn}), and $Z_{\br}$ is an $\rSU2$ matrix.  Really, $Z_{\br}$ is a set of operators giving the \emph{position coordinates} of a particle moving in the space of $\rSU2$ matrices.  There are also corresponding angular momentum operators that generate right- and left- rotations of $Z_{\br}$.  
The representation Eq.~(\ref{eqn:splitting-Psi}) has a \emph{local} $\rSU2$ redundancy, and, upon enforcing an appropriate local constraint, the Hubbard model can be exactly rewritten as an $\rSU2$ gauge theory.

While the development in Sec.~\ref{sec:su2-formulation} is specific to the Hubbard model on a bipartite lattice, it is straightforward to apply this formulation to a general Hubbard model with only minor modifications.  This is sketched in Sec.~\ref{sec:frust-lattice}; further development of this case is left to future work.

In Sec.~\ref{sec:mft}, a mean-field theory is constructed to describe spin liquid and valence-bond solid insulating phases.  The mean-field theory can also access the Mott transition from a given spin liquid phase to a corresponding conducting ground state.  We have considered a restricted class of mean-field ansatz, in order to focus on the simplest possibility of a direct, continuous transition from a spin liquid to the semimetal; in Sec.~\ref{sec:mft-phasediag} the mean-field phase diagram is discussed.  The saddle point corresponding to the $\rSU2$ ASL is found to be the lowest energy state near the Mott transition, and the stability of this state is enhanced by the addition of the third-neighbor hopping $t''$.

In Sec.~\ref{sec:effective-theory} we develop an effective low-energy theory that encapsulates the universal features of the $\rSU2$ ASL, as well as those of the Mott transition and the conducting state (the semimetal) on the other side.  The low-energy theory consists of fermionic spinons and rotor bosons coupled to a fluctuating, emergent $\rSU2$ gauge field.  The continuum field theory of the spin liquid consists of Dirac fermions coupled to the $\rSU2$ gauge field, and from this starting point we find enhanced N\'{e}el \emph{and} VBS correlations in the spin liquid state.  This leads to specific suggestions for the detection of the spin liquid in numerical simulations (Sec.~\ref{sec:spin-liquid-ft}).
The field theory of the Mott transition, which also includes a charge-carrying bosonic matter field in the fundamental representation of the $\rSU2$ gauge group, is written down and briefly discussed in Sec.~\ref{sec:mott-trans-ft}.  In Sec.~\ref{sec:u1sl}, the connection between pseudospin breaking and the $\rU1$ slave-rotor formulation is made more concrete.  At the level of the effective theory, it is shown that 
pseudospin breaking of the type induced by ${\cal H}_V$ can lead to a phase transition from the $\rSU2$ ASL to the $\rU1$ ASL.  This spin liquid state, as well as its own direct transition to the semimetal, are described by the effective $\rU1$ gauge theory of Ref.~\onlinecite{sslee05}, which therefore applies here in the presence of strong enough pseudospin breaking.  

In Sec.~\ref{sec:discussion} we summarize our main results for the honeycomb lattice model, with a focus on implications for future numerical simulations.  Furthermore, we briefly discuss the possibilities for applying the $\rSU2$ slave-rotor formulation to more general Hubbard models, and also speculate on its possible utility in obtaining a better physical picture of fermionic spinons.  Various technical details are described in the appendices.

\section{Relation to earlier work}
\label{sec:earlier-work}
The $\rSU2$ slave-rotor formulation has recently been constructed independently by K.-S. Kim via a functional integral approach.\cite{kskim06a,kskim06b}  Refs.~\onlinecite{kskim06a,kskim06b} began with the square lattice Hubbard model and decoupled the on-site repulsive interaction with a pseudospin vector field.  Then, after a series of changes of variables, the microscopic $\rSU2$ slave-rotor functional integral of Eqs.~(\ref{eqn:sz-micro}-\ref{eqn:st-micro}) was obtained.  Here, in contrast, we introduce the $\rSU2$ slave-rotor variables working in the Hamiltonian, and then go to a functional integral.  Our approach makes clear that this formulation is a generalization of the $\rSU2$ gauge theory of the Heisenberg model to the case of the Hubbard model.

Despite the equivalence of our formulation to that of Kim, we believe that this formulation has been analyzed incorrectly in Refs.~\onlinecite{kskim06a,kskim06b}.  Upon moving away from half-filling, Refs.~\onlinecite{kskim06a,kskim06b} found a non-Fermi liquid metallic phase intervening between a conventional Fermi liquid and a spin liquid state.  First, the spin liquid state certainly cannot persist away from half-filling; the rotor bosons will enter the ground state at finite density and condense, resulting in a conducting ground state (either a Fermi liquid metal or a superconductor) with the gauge field in the Higgs phase.  Second, the $\rSU2$ slave-rotor formulation probably cannot be used to describe a non-Fermi liquid metallic phase.  We believe the latter error arises because Refs.~\onlinecite{kskim06a, kskim06b} do not correctly take into account the separate roles of pseudospin symmetry and $\rSU2$ gauge redundancy.  We comment on this in more detail at the end of Sec.~\ref{sec:basic-effective-theory}.

We also mention earlier work by Wu \emph{et. al.}, where the idea of quantum disordering a $d$-wave superconductor by working with the ${\rm O}(3)$-vector pseudospin order parameter was explored.\cite{wu98}  These ideas are connected to the approach discussed here.

\section{Review of pseudospin}
\label{sec:pseudospin}

Here we briefly review the pseudospin symmetry of Hubbard models on bipartite lattices.\cite{cnyang89,zhang90}  While pseudospin symmetry will not be present (except perhaps approximately) in any realistic Hubbard model, even if the lattice is bipartite, it is still important to consider it for the analysis in this paper.  First, it is present for the simplest Hubbard models on bipartite lattices, namely for Hamiltonians with only bipartite hopping and on-site repulsion.  Because we propose to study the honeycomb Hubbard model via quantum Monte Carlo simulation, it is important to take pseudospin into account.  Second, it will turn out to lead us to a generalization of the $\rSU2$ gauge theory of the Heisenberg model to the case of the Hubbard model, which will be useful much more generally.

The most familiar symmetries of the Hubbard Hamiltonian Eq.~(\ref{eqn:hubbard-model}), in addition to particle-hole symmetry, are the space group symmetry, time reversal, $\rU1$ charge rotations generated by the total electron number, and $\rSU2$ spin rotation generated by $\bS_{{\rm tot}} \equiv \sum_{\br} \bS_{\br}$, where
$\bS_{\br} = c^\dagger_{\br \alpha} \boldsymbol{\sigma}_{\alpha \beta} c^{\vphantom\dagger}_{\br \beta} /2$ and $\boldsymbol{\sigma} = (\sigma^1, \sigma^2, \sigma^3)$ is a vector of the $2 \times 2$ Pauli matrices.  The pseudospin symmetry is most easily exposed by organizing the electron operators into the $2 \times 2$ matrix $\Psi_{\br}$ as in Eq.~(\ref{eqn:Psi-defn}).  Left-$\rSU2$ rotations of $\Psi_{\br}$ are generated by the spin density:
\begin{equation}
e^{i \balpha \cdot \bS_{\br} } \Psi_{\br} e^{-i \balpha \cdot \bS_{\br} } = e^{ - i \balpha \cdot \boldsymbol{\sigma} / 2} \Psi_{\br} \text{.}
\end{equation}

The pseudospin density $\bT_{\br}$ is a spin singlet and  is defined as 
\begin{eqnarray}
T^z_{\br} &=& \frac{1}{2}(1 - c^\dagger_{\br \alpha} c^{\vphantom\dagger}_{\br \alpha} ) \label{eqn:tz-defn} \\
T^{+}_{\br} &=& \er c_{\br \uparrow} c_{\br \downarrow} \\
T^{-}_{\br} &=& \er c^\dagger_{\br \downarrow} c^\dagger_{\br \uparrow} \text{,}
\end{eqnarray}
where $T^{\pm}_{\br} \equiv T^x_{\br} \pm i T^y_{\br}$, and
\begin{equation}
\er = \left\{\begin{array}{ll}1, & \br \in A \\ -1, & \br \in B \end{array} \right. \text{.}
\end{equation}
The pseudospin density generates right-$\rSU2$ rotations of $\Psi_{\br}$:
\begin{equation}
e^{i \balpha \cdot \bT_{\br} } \Psi_{\br} e^{-i \balpha \cdot \bT_{\br} } = \Psi_{\br} e^{ i \balpha \cdot \boldsymbol{\sigma} / 2}  \text{.}
\end{equation}

The Hubbard Hamiltonian is manifestly invariant both under spin and pseudospin symmetry in the form
\begin{equation}
\label{eqn:hubbard-T-form}
{\cal H}_{{\rm H}} = - \sum_{( \br, \br' )} t_{\br \br'} \tr (\Psi^\dagger_{\br} \Psi_{\br'} )
+ \frac{2 U}{3} \sum_{\br} \big( \bT^2_{\br} - \frac{3}{8} \big) \text{,}
\end{equation}
where the first term is already Hermitian without taking the complex conjugate.  It should be noted that the bipartite nature of the lattice was crucial in obtaining this form; it entered via the factors of $\epsilon_{\br}$ in the definitions of $\Psi_{\br}$ and $\bT_{\br}$.
It is also useful to note that spin and pseudospin densities can be expressed simply in terms of $\Psi_{\br}$:
\begin{eqnarray}
\bS_{\br} &=& \frac{1}{4} \tr (\Psi^\dagger_{\br} \boldsymbol{\sigma} \Psi_{\br} ) \\
\bT_{\br} &=& \frac{1}{4} \tr ( \Psi^{\vphantom\dagger}_{\br} \bsigma \Psi^\dagger_{\br} ) \text{.}
\end{eqnarray}

Pseudospin is an enlargement of the ${\rm O}(2)$ symmetry group generated by $\rU1$ charge rotations and particle-hole symmetry.  Under a $\rU1$ charge rotation $c_{\br \alpha} \to e^{-i \theta} c_{\br \alpha}$ and hence $\Psi_{\br} \to \Psi_{\br} e^{- i \theta \sigma^3}$, so the conserved $\rU1$ charge density is simply $- 2 T^z_{\br}$, as is apparent from Eq.~(\ref{eqn:tz-defn}).  Particle-hole symmetry acts on the electron operators as $c_{\br \alpha} \to \er (i \sigma^2)_{\alpha \beta} c^\dagger_{\br \beta}$, and so $\Psi_{\br} \to \Psi_{\br} (-i \sigma^2)$.  Particle-hole symmetry is therefore implemented by the unitary operator $\exp \big( - \frac{i \pi}{2} \sum_{\br} T^y_{\br} \big)$, which is again part of the pseudospin symmetry group.

To understand a little better the physical meaning of pseudospin symmetry, it is useful to consider the \emph{staggered} pseudospin $\bT^s_{\br} = \er \bT_{\br}$.  The $z$-component $(\bT^s)^z$ is the order parameter for a staggered charge density wave state, while the transverse components make up the order parameter for an $s$-wave superconductor.  The pseudospin symmetry rotates among these superficially unrelated states.

Pseudospin is broken by the nearest-neighbor repulsion term ${\cal H}_V$ of Eq.~(\ref{eqn:hv-term}); in terms of pseudospin generators ${\cal H}_V$ can be written
\begin{equation}
{\cal H}_V = 4 V \sum_{\langle \br \br' \rangle} T^z_{\br} T^z_{\br'} \text{.}
\end{equation}
This is an \emph{Ising} exchange in pseudospin space.  It breaks the pseudospin symmetry down to the ${\rm O}(2)$ subgroup generated by $\rU1$ charge rotations and particle-hole symmetry.

\section{Slave-particle formulation}
\label{sec:slavep}

Here, we develop the $\rSU2$ slave-rotor formulation.  We are led to this formulation in the context of Hubbard models on bipartite lattices (\emph{e.g.} the honeycomb lattice), as a means to properly account for the pseudospin symmetry.  However, its real power is as an extension of the $\rSU2$ gauge theory of the Heisenberg model to the case of the Hubbard model.  As such, it should be useful much more generally, even in the absence of pseudospin symmetry.

First, in Sec.~\ref{sec:u1-attempt}, we review the $\rU1$ slave-rotor formulation,\cite{florens04,sslee05} and explain why it is not adequate to describe the honeycomb lattice model.  Next we build on the $\rU1$ formulation and develop the $\rSU2$ slave-rotor formulation in the context of a Hubbard model on a bipartite lattice (Sec.~\ref{sec:su2-formulation}).  Finally, in Sec.~\ref{sec:frust-lattice} we sketch the development of the $\rSU2$ formulation for a general Hubbard model, without the assumption of a bipartite lattice.  The remainder of the paper is devoted to an analysis of the bipartite case and specifically the honeycomb lattice model.  Applications of the general formalism are left to future work.

\subsection{Attempt at a $\rU1$ formulation}
\label{sec:u1-attempt}

The $\rU1$ formulation begins by rewriting the electron as the product of a charge-neutral fermionic spinon and a charge-carrying ${\rm O}(2)$ rotor
\begin{equation}
\label{eqn:u1-splitting}
c_{\br \alpha} = e^{-i \phi_{\br}} f_{\br \alpha} \text{.}
\end{equation}
This is a rewriting of the electron in terms of spin-charge separated variables.  In Eq.~(\ref{eqn:u1-splitting}), $f_{\br \alpha}$ is the spinon destruction operator, and the operator $e^{i \phi_{\br}}$ measures the position coordinate of the ${\rm O}(2)$ rotor, which is constrained to lie on the unit circle in the complex plane.  There is also a conjugate rotor angular momentum $L_{\br}$ taking integer eigenvalues, and satisfying the commutation relation 
$[ L_{\br} , e^{\pm i \phi_{\br'}} ]
= \pm \delta_{\br \br'} e^{\pm i \phi_{\br}}$.  The Hilbert space on each site is a product of spinon and rotor Hilbert spaces, and a local constraint must be imposed to restrict this to the physical Hilbert space of the original Hubbard model.  This is accomplished by requiring
\begin{equation}
\label{eqn:u1-constraint}
f^\dagger_{\br \alpha} f^{\vphantom\dagger}_{\br \alpha} - L_{\br} = 1 \text{.}
\end{equation}
In these variables there is a local $\rU1$ redundancy, where all physical operators are invariant under the transformation $f_{\br \alpha} \to e^{i \lambda_{\br}} f_{\br \alpha}$ and $\phi_{\br} \to \phi_{\br} + \lambda_{\br}$.

As it stands, Eqs.~(\ref{eqn:u1-splitting}) and~(\ref{eqn:u1-constraint}) can be used to exactly rewrite the Hubbard Hamiltonian as a $\rU1$ gauge theory, but this does not yet provide any information about the low-energy physics.  To proceed, one can develop a mean-field theory in terms of the fermions and rotors and expand in the fluctuations about the mean field saddle point.  This was done for the honeycomb Hubbard model in Ref.~\onlinecite{sslee05}; the result can be encapsulated in a Hamiltonian $\rU1$ lattice gauge theory, and in the mean-field analysis of the honeycomb Hubbard model a $\rU1$ algebraic spin liquid was found adjacent to the Mott transition.  In the low-energy Hamiltonian, the local constraint is modified to
\begin{equation}
(\operatorname{div} e)_{\br} = f^\dagger_{\br \alpha} f^{\vphantom\dagger}_{\br \alpha} - L_{\br} - 1 \text{,}
\end{equation}
where $e_{\br \br'}$ is the integer-valued emergent electric field lying on the links of the honeycomb lattice, and $(\operatorname{div} e)_{\br} = \sum_{\br' \text{ n.n. } \br} e_{\br \br'}$ is its lattice divergence.  There is a term in the effective Hamiltonian of the form $h \sum_{\langle \br \br' \rangle} e^2_{\br \br'}$, so, when $h < \infty$, $\langle e^2_{\br \br'} \rangle \neq 0$ and the local constraint has been \emph{softened}.  For $h = \infty$ and a hard constraint, the only states of the rotor angular momentum allowed are $L = 0, \pm 1$.  However, in order to construct a useful effective theory the constraint must be softened, and higher angular momentum states will be present.

In terms of pseudospin, the $L = 0$ state is a pseudospin singlet, and the $L = \pm 1$ states form a pseudospin doublet.  But the higher angular momentum states do not transform as irreducible representations of the pseudospin, and as soon as they are present, the resulting effective Hamiltonian  explicitly breaks the pseudospin symmetry.  The $\rU1$ formulation therefore leads to effective theories that do not respect the pseudospin symmetry; this invalidates its results, at least near the metal-insulator transition.

If the pseudospin symmetry is broken explicitly in the microscopic Hamiltonian -- for example, by adding the term ${\cal H}_V$ of Eq.~(\ref{eqn:hv-term}) -- the $\rU1$ formulation remains valid, in the sense that it leads to a sensible low-energy theory respecting the symmetries of the model.  The low-energy theories obtained in the $\rU1$ formulation can be invariant under the ${\rm O}(2)$ subgroup of the pseudospin generated by $\rU1$ charge rotations and particle-hole symmetry, so the formulation can be used in models with these symmetries, or with less symmetry (as in the triangular lattice Hubbard model, which was the focus of Ref.~\onlinecite{sslee05}).\footnote{It should also be straightforward to modify the $\rU1$ formulation to apply when the pseudospin symmetry has been \emph{spontaneously} broken, if there is a motivation to do so.}

In fact, we can work backwards from the effective $\rU1$ lattice gauge theory Hamiltonian to obtain a modified Hubbard model, which itself breaks pseudospin symmetry.  This is done by taking $h$ large compared to all other energies in the effective Hamiltonian, and doing degenerate perturbation theory in the manifold of states where $e_{\br \br'}  = 0$.  One term that will be present is a spinon hopping term with coefficient $t_s$.  At order $t_s^2 / h$ in perturbation theory (the leading order), two terms are generated.  The first is an antiferromagnetic Heisenberg exchange.  The second is nothing but the repulsive interaction ${\cal H}_V$, with $V \propto t_s^2 /h$.  The interaction so obtained is repulsive, and it is natural to guess that positive (rather than negative) $V$ is more likely to favor a $\rU1$ spin liquid state.

\subsection{$\rSU2$ slave-rotor formulation and spin-pseudospin separation}
\label{sec:su2-formulation}

The discussion above shows that the $\rU1$ formulation needs to be generalized to respect the pseudospin symmetry.  To accomplish this, we need to generalize spin-charge separation to \emph{spin-pseudospin} separation -- recall that the charge-rotation symmetry is a subgroup of the pseudospin.  As is apparent from the $\Psi_{\br}$ matrix of Eq.~(\ref{eqn:Psi-defn}), the electron operator transforms as a doublet under both spin and pseudospin.  Therefore we should split it into a doublet of fermionic spinons, and, in contrast to the $\rU1$ formulation, a doublet of bosonic objects carrying the pseudospin.  As in the $\rU1$ formulation, we shall take this bosonic object to be a kind of rotor.

We proceed in this manner to develop the $\rSU2$ slave-rotor formulation.
  We begin by writing
\begin{equation}
\label{eqn:Psi-split}
\Psi_{\br} = F_{\br} Z_{\br} \text{,}
\end{equation}
where
\begin{equation}
F_{\br} = \left(
\begin{array}{cc}
f_{\br \uparrow} & \er f^\dagger_{\br \downarrow} \\
f_{\br \downarrow} & -\er f^{\dagger}_{\br \uparrow}
\end{array}\right)
\end{equation}
is a matrix of $S = 1/2$ spinon creation and destruction operators with anticommutation relations
$\{ f_{\br \alpha} , f^\dagger_{\br' \beta} \} = \delta_{\alpha \beta} \delta_{\br \br'}$, and $Z_{\br}$ is an $\rSU2$ matrix.  Actually $Z_{\br}$ is a $2 \times 2$ matrix of operators measuring the position coordinate of an $\rSU2$-matrix rotor, which can be thought of as a quantum particle constrained to move in the space of $\rSU2$ matrices.  
The matrix $Z_{\br}$ can be parametrized by writing
\begin{equation}
\label{eqn:z-parametrization}
Z_{\br} = \left(
\begin{array}{cc}
z_{\br 1} & z_{\br 2} \\
-z^*_{\br 2} & z^*_{\br 1} 
\end{array} \right) \text{,}
\end{equation}
where the $z_{\br i}$ are operators,  $z^*_{\br i} \equiv z^\dagger_{\br i}$, and there is a constraint $|z_{\br 1}|^2 + |z_{\br 2}|^2 = 1$.  This parametrization makes clear that the $\rSU2$ matrix rotor is \emph{identical} to the more familiar ${\rm O}(4)$ rotor, a quantum particle constrained to the 3-sphere.   Furthermore, it is important to note that the decomposition Eq.~(\ref{eqn:Psi-split}) contains the minimal number of degrees of freedom for spin-pseudospin separation:  a \emph{single} doublet of spinons, and a \emph{single} doublet of pseudospin-carrying rotor bosons.

  There is a \emph{local} $\rSU2$ redundancy, and a corresponding invariance under the transformation 
$F_{\br} \to F_{\br} e^{-i \alpha_{\br} \cdot \bsigma /2}$ and $Z_{\br} \to e^{i \balpha_{\br} \cdot \bsigma / 2} Z_{\br}$.  This will allow us to rewrite the Hubbard model exactly as an
 $\rSU2$ gauge theory.  In fact, focusing only on the spinons, as is appropriate in the $U \to \infty$ limit, the $\rSU2$ gauge structure reduces to the one present in the Heisenberg model.\cite{affleck88a,dagotto88}  This was extended to to the case of a doped system described by the $t$-$J$ model in Refs.~\onlinecite{wen96} and~\onlinecite{palee98}.  In the present work, we show that this $\rSU2$ gauge structure extends to the full Hubbard model.

The enlarged slave-particle Hilbert space on each site consists of a product of the Hilbert space of $S = 1/2$ fermionic spinons and that of the $\rSU2$ matrix rotor.  To understand this in more detail we first state some facts about the $\rSU2$ matrix rotor that are explained more completely in Appendix~\ref{app:su2rotor}.  The Hilbert  space for a single $\rSU2$ matrix rotor has a basis of position eigenstates $\{ | Z \rangle \}$ satisfying $\hat{Z} | Z \rangle = Z | Z \rangle$; here, $Z \in \rSU2$ and $\hat{Z}$ is the $2 \times 2$ matrix of operators discussed above, and we use the ``hat'' symbol as a reminder that an object is an operator when it is not otherwise clear from the context.  We define the analog of angular momentum operators $\bJ_R$ and $\bJ_L$ as the generators of right- and left-$\rSU2$ rotations:
\begin{eqnarray}
e^{i \balpha \cdot \bJ_R} | Z \rangle &=& | Z e^{- i \balpha \cdot \bsigma / 2} \rangle \text{,} \\
e^{i \balpha \cdot \bJ_L} | Z \rangle &=& | e^{i \balpha \cdot \bsigma / 2} Z \rangle \text{.}
\end{eqnarray}
This implies the operator identities
\begin{eqnarray}
e^{i \balpha \cdot \bJ_R } \hat{Z} e^{-i \balpha \cdot \bJ_R } &=& \hat{Z} e^{i \balpha \cdot \bsigma / 2} \text{,} \\
e^{i \balpha \cdot \bJ_L } \hat{Z} e^{-i \balpha \cdot \bJ_L } &=& e^{- i \balpha \cdot \bsigma / 2} \hat{Z} \text{,}
\end{eqnarray}
and the commutation relations
\begin{eqnarray}
{[} J^i_R , J^j_R  {]} &=& i \epsilon^{i j k} J^k_R  \\
{[} J^i_L , J^j_L  {]} &=& i \epsilon^{i j k} J^k_L \\
{[} J^i_R , J^j_L  {]} &=& 0 \\
{[} J^i_R , \hat{Z} {]} &=& \hat{Z} \Big( \frac{\sigma^i}{2} \Big) \\
{[} J^i_L , \hat{Z} {]} &=& - \Big( \frac{\sigma^i}{2} \Big) \hat{Z} \text{.}
\end{eqnarray}

We can also work with the basis of angular momentum eigenstates specified by $\{ | \ell_R = \ell, \ell_L = \ell, m_R, m_L \rangle \}$, where $\ell_R$ and $\ell_L$ are the total right and left angular momenta, respectively, and $m_R$ and $m_L$ are the corresponding projections along the $z$-axis.  It is shown in Appendix~\ref{app:su2rotor} that only states with $\ell_R = \ell_L$ are allowed in the Hilbert space; this is a higher-dimensional analog of the more familiar fact that, for a particle moving on the 2-sphere, only integer angular momenta are allowed.  The unique rotationally invariant state is denoted $| 0 \rangle_{{\rm rot}} \equiv |0,0,0,0 \rangle$.

Now we can return to the Hubbard model, where we place an $\rSU2$ matrix rotor on every site.  The angular momenta are denoted
by $\bJ_R(\br)$ and $\bJ_L(\br)$.   The pseudospin is clearly identified with $\bJ_R(\br)$:
\begin{equation}
\bT_{\br} = \bJ_R(\br) \text{.}
\end{equation}
We now have enough information to represent the physical states for one site $\br$ of the Hubbard in the enlarged slave-particle Hilbert space.  The physical vacuum state for the electrons (the empty state) is denoted by $| 0 \rangle$, and, in the slave-particle Hilbert space, we define a vacuum state $|0 \rangle_{{\rm sp}} = |0 \rangle_f \otimes |0\rangle_{{\rm rot}}$, where $|0\rangle_f$ is the spinon vacuum.  Then we have
\begin{eqnarray}
c^\dagger_{\br \uparrow} |0\rangle &\leftrightarrow&  f^\dagger_{\br \uparrow} | 0\rangle_{{\rm sp}} \\
c^\dagger_{\br \downarrow} |0\rangle &\leftrightarrow&  f^\dagger_{\br \downarrow} | 0\rangle_{{\rm sp}} \\
|0\rangle &\leftrightarrow& ( z_{\br 1} + \er z^*_{\br 2} f^\dagger_{\br \uparrow} f^\dagger_{\br \downarrow} ) | 0 \rangle_{{\rm sp}} \\
c^\dagger_{\br \uparrow} c^\dagger_{\br \downarrow}|0\rangle &\leftrightarrow&
(z^*_{\br 1} f^\dagger_{\br \uparrow} f^\dagger_{\br \downarrow} - \er z_{\br 2} ) | 0 \rangle_{{\rm sp}} \text{.}
\end{eqnarray}
It can be shown that the electron operator in the slave-particle Hilbert space, as defined by Eq.~(\ref{eqn:Psi-split}), has the correct matrix elements among all these states, as well as the correct anti-commutation relations.

We still need a local constraint in order to eliminate the unphysical states in the slave-particle Hilbert space.  As in other slave-particle gauge theories, the appropriate constraint is related to the $\rSU2$ redundancy; in fact, it involves the generator of $\rSU2$ gauge transformations.  We define
\begin{equation}
\label{eqn:JF-expression}
\bJ_F(\br) =  \frac{1}{4} \tr (F_{\br} \bsigma F^\dagger_{\br} ) \text{,}
\end{equation}
which generates right-$\rSU2$ rotations of $F_{\br}$,
and
\begin{equation}
\bJ_G(\br) \equiv \bJ_F(\br) + \bJ_L(\br) \text{.}
\end{equation}
We have
\begin{eqnarray}
e^{i \balpha \cdot \bJ_G(\br) } Z_{\br} e^{-i \balpha \cdot \bJ_G(\br) } &=& e^{- i \balpha \cdot \bsigma / 2} Z_{\br} \\
e^{i \balpha \cdot \bJ_G(\br) } F_{\br} e^{-i \balpha \cdot \bJ_G(\br) } &=& F_{\br} e^{ i \balpha \cdot \bsigma / 2} \text{,}
\end{eqnarray}
so that $\bJ_G(\br)$ is the generator of $\rSU2$ gauge transformations; equivalently, it is the $\rSU2$ gauge charge.  It can be shown that, under the local constraint
\begin{equation}
\bJ_G(\br) = 0 \text{,}
\end{equation}
the slave-particle Hilbert space is identical to the physical Hilbert space of the Hubbard model.

The relationship of this formulation to the $\rSU2$ gauge theory of the Heisenberg model is more transparent if we introduce the spinor
\begin{equation}
\label{eqn:psi-defn}
\psi_{\br} = \left( \begin{array}{c}
f^\dagger_{\br \uparrow} \\
\er f_{\br \downarrow}
\end{array} \right) \text{,}
\end{equation}
which transforms as a doublet under $\rSU2$ gauge rotation.  In terms of $\psi_{\br}$ the fermion part of the $\rSU2$ gauge charge takes the more familiar form
\begin{equation}
\bJ_F(\br) = \frac{1}{2} \psi^\dagger_{\br} \bsigma \psi^{\vphantom\dagger}_{\br} \text{.}
\end{equation}

Now we can construct the functional integral representation of the Hubbard model in the $\rSU2$ slave rotor formulation, using the results above, and those of Appendix~\ref{app:su2rotor}.  Dropping overall additive constants, we write the Hamiltonian as
\begin{equation}
{\cal H}_{{\rm H}} = - \sum_{( \br, \br' )} t_{\br \br'} \tr \big[ ( F_{\br} Z_{\br} )^\dagger (F_{\br'} Z_{\br'}) \big]
+ \frac{2 U}{3} \sum_{\br} \bT_{\br}^2 \text{,}
\end{equation}
with the constraint
\begin{equation}
\bJ_F(\br) + \bJ_L(\br) = 0 \text{.}
\end{equation}
We can follow the methods of Appendix~\ref{app:su2rotor} to write down the functional integral
\begin{equation}
{\cal Z} = \int {\cal D}Z {\cal D}F {\cal D} \boldsymbol{a}_\tau e^{- S} \text{,}
\end{equation}
where $S = S_Z + S_F + S_t $.  We concentrate on zero-temperature; the imaginary time $\tau$ ranges from $-\infty$~to~$\infty$ and we write $\int d\tau \equiv \int^{\infty}_{-\infty} d\tau$.
We have
\begin{equation}
\label{eqn:sz-micro}
S_Z = \frac{3}{4 U} \sum_{\br} \int d\tau \tr \Big[
Z^\dagger_{\br} \big(\overleftarrow{\partial}_\tau - \frac{i \ba_\tau \cdot \bsigma}{2} \big)
\big(\partial_\tau + \frac{i \ba_\tau \cdot \bsigma}{2} \big) Z_{\br} \Big] \text{,}
\end{equation}
where the symbol $\overleftarrow{\partial}_{\tau}$ is a derivative \emph{acting on the left}.  Also,
\begin{eqnarray}
S_F &=& \sum_{\br} \int  d\tau \Big[ \bar{f}_{\br \alpha} \partial_\tau f_{\br \alpha}
+ i \ba_\tau \cdot \bJ_F(\br, \tau) \Big] \nonumber \\
&=& \sum_{\br} \int  d\tau\, \psi^\dagger_{\br} ( \partial_\tau + \frac{i \ba_\tau \cdot \bsigma}{2} )
\psi_{\br}  \nonumber \\
&=& \frac{1}{2} \sum_{\br} \int  d\tau\, \tr \Big[ F_{\br} ( \partial_\tau + \frac{i \ba_\tau \cdot \bsigma}{2} ) F^\dagger_{\br} \Big] \label{eqn:sf-micro} \text{,}
\end{eqnarray}
and
\begin{equation}
S_t = - \int  d\tau \sum_{( \br, \br' )}  t_{\br \br'} \tr \big( (F_{\br} Z_{\br} )^\dagger (F_{\br'} Z_{\br'}) \big) \text{.}
\label{eqn:st-micro}
\end{equation}

So far, this is an exact rewriting of the Hubbard model as a (strong coupling) $\rSU2$ gauge theory.  Both spin and pseudospin symmetry are manifest in this representation.  A general gauge transformation acts as follows:
\begin{eqnarray}
Z_{\br}(\tau) &\to& G^\dagger_{\br}(\tau) Z_{\br}(\tau) \\
F_{\br}(\tau) &\to& F_{\br}(\tau) G_{\br}(\tau) \\
\frac{i \ba_\tau \cdot \bsigma}{2} &\to& 
G^\dagger_{\br}(\tau) \Big( \frac{i \ba_\tau \cdot \bsigma}{2}  + \partial_\tau \Big) G_{\br}(\tau) \text{,}
\end{eqnarray}
where $G_{\br}(\tau)$ is an $\rSU2$ matrix.

\subsection{$\rSU2$ slave-rotor formulation (general lattice)}
\label{sec:frust-lattice}

The $\rSU2$ slave-rotor formulation is really a generalization of the $\rSU2$ gauge theory of the Heisenberg model to \emph{any} Hubbard model.  In particular, pseudospin symmetry is not required, even though we have been focussing on models that possess it up to now.  Here, we digress from the honeycomb lattice problem and develop the $\rSU2$ slave-rotor formalism for a general Hubbard model, without the assumption of a bipartite lattice and pseudospin symmetry; we also include an arbitrary chemical potential.  It turns out that only very simple modifications are required.  Because we shall not return to this more general case later in the paper, it is most economical to proceed by changing notation.  This section is therefore meant to be self-contained, and \emph{the notation here is not consistent with the rest of the paper}.

We organize the electron operators into the the $2 \times 2$ matrix
\begin{equation}
\Psi_{\br} = \left( \begin{array}{cc}
c_{\br \uparrow} & c^\dagger_{\br \downarrow} \\
c_{\br \downarrow} & - c^\dagger_{\br \uparrow}
\end{array} \right) \text{.}
\end{equation}
The difference from Eq.~(\ref{eqn:Psi-defn}) is the absence of $\epsilon_{\br}$, which is not defined for a general lattice.  Spin rotations act on $\Psi_{\br}$ as before
\begin{equation}
e^{i \balpha \cdot \bS_{\br} } \Psi_{\br} e^{-i \balpha \cdot \bS_{\br} } = e^{ - i \balpha \cdot \boldsymbol{\sigma} / 2} \Psi_{\br} \text{,}
\end{equation}
and, also as before, $\rU1$ charge rotations act by sending $\Psi_{\br} \to \Psi_{\br} e^{-i \theta \sigma^3}$.  Even though pseudospin is not in general a symmetry, it is useful to note that the generators of right-$\rSU2$ rotations of $\Psi_{\br}$ are $\bT_{\br}$, 
so that,
\begin{equation}
e^{i \balpha \cdot \bT_{\br} } \Psi_{\br} e^{-i \balpha \cdot \bT_{\br}} = \Psi_{\br} e^{i \balpha \cdot \bsigma / 2} \text{,}
\end{equation}
where
\begin{equation}
T^z_{\br} = \frac{1}{2} \big( 1 - c^\dagger_{\br \alpha} c^{\vphantom\dagger}_{\br \alpha} \big) \text{,}
\end{equation}
and
\begin{equation}
T^+_{\br} = c_{\br \uparrow} c_{\br \downarrow} \text{.}
\end{equation}
Also, $T^{-}_{\br} = (T^+_{\br})^\dagger$ and $T^{\pm}_{\br} = T^x_{\br} \pm i T^y_{\br}$.  Dropping additive constants and including a chemical potential $\mu$, the Hubbard Hamiltonian can now be written
\begin{equation}
{\cal H}_{{\rm H}} = - \sum_{( \br, \br' )} t_{\br \br'} \tr ( \sigma^3 \Psi^\dagger_{\br} \Psi_{\br'} )
+ \sum_{\br} \Big( \frac{2 U}{3} \bT^2_{\br} + 2 \mu T^z_{\br} \Big) \text{,}
\end{equation}
where the difference from Eq.~(\ref{eqn:hubbard-T-form}) is the presence of the $\sigma^3$ matrix in the hopping term.

We make the decomposition
\begin{equation}
\Psi_{\br} = F_{\br} Z_{\br} \text{,}
\end{equation}
where
\begin{equation}
\label{eqn:general-F}
F_{\br} = \left( \begin{array}{cc}
f_{\br \uparrow} & f^\dagger_{\br \downarrow} \\
f_{\br \downarrow} & - f^\dagger_{\br \uparrow}
\end{array} \right) \text{,}
\end{equation}
and $Z_{\br}$ is defined as before.  $\Psi_{\br}$ is invariant under the $\rSU2$ gauge transformation 
$F_{\br} \to F_{\br} e^{ i \balpha_{\br} \cdot \bsigma}$ and $Z_{\br} \to e^{- i \balpha_{\br} \cdot \bsigma} Z_{\br}$. 
Again, we define the generators of right and left rotations of $Z_{\br}$ to be $\bJ_R(\br)$ and $\bJ_L(\br)$, respectively, and we identify $\bJ_R(\br) = \bT_{\br}$.  Defining
\begin{equation}
\bJ_F(\br) = \frac{1}{4} \tr \big( F_{\br} \bsigma F^\dagger_{\br} \big) \text{,}
\end{equation}
the $\rSU2$ gauge charge is 
\begin{equation}
\bJ_G(\br) = \bJ_F(\br) + \bJ_L(\br) \text{.}
\end{equation}
The local constraint $\bJ_G(\br) = 0$ restricts the slave-particle Hilbert space to the physical Hilbert space of the Hubbard model.

Following the methods of Appendix~\ref{app:su2rotor}, we can write down the functional integral.  The partition function is
\begin{equation}
{\cal Z} = \int {\cal D}Z {\cal D}F {\cal D} \boldsymbol{a}_\tau e^{- S} \text{,}
\end{equation}
where $S = S_Z + S_F + S_t$, $S_F$ is identical to Eq.~(\ref{eqn:sf-micro}) upon replacing $F_{\br}$ there by the definition Eq.~(\ref{eqn:general-F}), and
\begin{widetext}
\begin{equation}
S_Z =  \frac{3}{4 U} \sum_{\br} \int d\tau \tr \Big[
Z^\dagger_{\br} \big(\overleftarrow{\partial}_\tau - \frac{i \ba_\tau \cdot \bsigma}{2} \big)
\big(\partial_\tau + \frac{i \ba_\tau \cdot \bsigma}{2} \big) Z_{\br} \Big] 
+ \frac{3 \mu}{2 U} \sum_{\br} \int d\tau \tr \Big[ \sigma^3 Z^\dagger_{\br} \big( \partial_\tau + \frac{i \ba_\tau \cdot \bsigma}{2} \big) Z_{\br} \Big] \text{.}
\end{equation}
\end{widetext}
Finally, we have
\begin{equation}
S_t = - \int  d\tau \sum_{( \br, \br' )}  t_{\br \br'} \tr \big( \sigma^3 (F_{\br} Z_{\br} )^\dagger (F_{\br'} Z_{\br'}) \big) \text{.}
\end{equation}

We have arrived at a representation of a general Hubbard model as an $\rSU2$ gauge theory.  The rest of this paper focuses on the formulation specific to bipartite lattices developed in Sec.~\ref{sec:su2-formulation}.  Using the following analysis as a guide, it should be possible in future work to use the more general formulation to investigate spin liquid physics near the Mott transition in a panoply of models.  Motivated by the experiments on $\kappa$-(BEDT-TTF)$_2$Cu$_2$(CN)$_3$, it would be particularly interesting to consider the triangular lattice Hubbard model and compare to the results of Ref.~\onlinecite{sslee05}.

\section{Mean-field theory}
\label{sec:mft}

We formulate a mean-field theory based on the gauge theory action for bipartite Hubbard models described above.   This will allow us to access spin-liquid Mott insulators, valence-bond solid states, and conducting states where the charge-carrying rotor bosons have condensed.  First, we develop the general structure of the mean-field theory (Sec.~\ref{mft-general-structure}).  Next, focusing on the honeycomb lattice, we consider a restricted class of mean-field ansatz and discuss the resulting phase diagram (Sec.~\ref{sec:mft-phasediag}).

\subsection{General structure of the mean-field theory}
\label{mft-general-structure}

The starting point for the mean-field theory is the functional integral derived in Sec.~\ref{sec:su2-formulation}.
We begin by trading in the constraint $Z_{\br}(\tau) \in \rSU2$ for a Lagrange multiplier.  The simplest way to do this is to parametrize $Z$ in terms of $z_1$ and $z_2$ as in Eq.~(\ref{eqn:z-parametrization}), allowing $z_1$ and $z_2$ to be arbitrary complex numbers, and then imposing the constraint $|z_1|^2 + |z_2|^2 = 1$ with a single Lagrange multiplier $\lambda_{\br}(\tau)$.  This results in a contribution to the action
\begin{equation}
S_\lambda = i \int  d\tau \sum_{\br} \lambda_{\br}(\tau) \Big[ \frac{1}{2} \tr (Z^\dagger_{\br} Z^{\vphantom\dagger}_{\br} ) - 1 \Big] \text{.}
\end{equation}

To decouple the hopping term $S_t$, we follow Ref.~\onlinecite{sslee05} and use the identity
\begin{equation}
e^{\epsilon \alpha_{i j} \beta_{i j} } = \frac{\epsilon}{\pi} \int d\eta_{i j} d\eta^*_{i j}
\exp \Big[ -\epsilon \big(
|\eta_{i j}|^2  - \eta_{i j} \alpha_{i j} - \eta_{i j}^* \beta_{i j} \big) \Big] \text{.}
\end{equation}
We write
\begin{equation}
S_t  = - \int  d\tau \sum_{(\br, \br')} t_{\br \br'} ( Z^{\vphantom\dagger}_{\br'} Z^\dagger_{\br} )_{\alpha \beta} (F^\dagger_{\br} F^{\vphantom\dagger}_{\br'})_{\beta \alpha} \text{.}
\end{equation}
Putting $\epsilon = | t_{\br \br'} | \Delta\tau$ and performing the decoupling using the complex fields $\eta^{\br \br'}_{\alpha \beta}$, we trade $S_t$ for the following contributions to the action:
\begin{eqnarray}
S_{\eta} &=& \int  d\tau \sum_{( \br, \br' )} |t_{\br \br'}| \tr \big( (\eta^{\br \br'})^\dagger \eta^{\br \br'} \big) \\
S_{tZ} &=& - \int  d\tau \sum_{( \br, \br' )} |t_{\br \br'}| \tr \big( Z^\dagger_{\br} \eta^{\br \br'} Z^{\vphantom\dagger}_{\br'} \big) \\
S_{tF} &=& \int  d\tau \sum_{( \br, \br' )} t_{\br \br'} \tr \big( F^{\vphantom\dagger}_{\br'} (\eta^{\br \br'})^\dagger F^\dagger_{\br} \big) \text{.}
\end{eqnarray}
By decoupling in this manner, we give each bond an orientation.  Because we are working with a bipartite lattice, it is natural to take $\br$ in the $A$ sublattice and $\br'$ in the $B$ sublattice in the equations above.  This orientation disappears in many of the mean-field saddle points discussed below.

We shall now parametrize $\eta^{\br \br'}$ in a more convenient form.  Because non-invertible matrices make up a measure-zero subset of all matrices, we may take $\eta^{\br \br'}$ to be invertible.  Then $\eta^{\br \br'}$ can be expressed as a complex number times an arbitrary matrix with unit determinant (such matrices form the group ${\rm SL}_2(\mathbb{C})$).  In fact, up to a set of measure zero, if $M \in {\rm SL}_2(\mathbb{C})$ then $M = \exp( i b_i \sigma^i)$ for some set of complex numbers $b_i$.\cite{hofmann78}
Therefore, we can write
\begin{equation}
\label{eqn:eta-param}
\eta^{\br \br'} = |\chi_{\br \br'}| e^{i \theta_{\br \br'}} \exp \Big[ i (c^{\br \br'}_{i} + i d^{\br \br'}_i) \sigma^i \Big] \text{,}
\end{equation}
where $|\chi_{\br \br'}|$, $\theta_{\br \br'}$, $c^{\br \br'}_i$ and $d^{\br \br'}_i$ are real numbers, and $|\chi_{\br \br'}| > 0$.

Eventually we will look for saddle points where we integrate out $Z_{\br}$ and $F_{\br}$ and work with a free energy depending on $\lambda$, $\ba_{\tau}$ and $\eta^{\br \br'}$.  We want to understand what kind of $\eta^{\br \br'}$ matrices give Hermitian $S_{tZ}$ and $S_{tF}$, and thus a real free energy.  Making use of Eq.~(\ref{eqn:psi-defn}) we have
\begin{equation}
\tr \big( F_{\br'} (\eta^{\br \br'})^\dagger F_{\br} \big) = \psi^\dagger_{\br'} (\eta^{\br \br'})^\dagger \psi^{\vphantom\dagger}_{\br}
+ \psi^\dagger_{\br} \sigma^2 (\eta^{\br \br'})^* \sigma^2 \psi^{\vphantom\dagger}_{\br} \text{.}
\end{equation}
Using Eq.~(\ref{eqn:eta-param}), we see that $S_{t F}$ is Hermitian if and only if $\theta_{\br \br'} = 0$ and $d^{\br \br'}_i = 0$, so that $\eta^{\br \br'}$ is a real number times an $\rSU2$ matrix.
To proceed similarly with $S_{tZ}$, we define
\begin{equation}
\zeta_{\br} \equiv 
\left(
\begin{array}{c} 
z_{\br 1} \\
-z^*_{\br 2}
\end{array} \right) \text{,}
\end{equation}
and we have
\begin{equation}
\tr \big( Z^\dagger_{\br} \eta^{\br \br'} Z_{\br'} \big) =
\zeta^\dagger_{\br}  \eta^{\br \br'} \zeta^{\vphantom\dagger}_{\br'}
+ \zeta^\dagger_{\br'} \sigma^2 (\eta^{\br \br'})^T \sigma^2 \zeta^{\vphantom\dagger}_{\br} \text{.}
\end{equation}
Again, this is Hermitian if and only if $\eta^{\br \br'}$ is a real number times an $\rSU2$ matrix.

We now write down the saddle-point equations:
\begin{eqnarray}
1 &=& \frac{1}{2} \left\langle \tr (Z^{\dagger}_{\br} Z^{\vphantom\dagger}_{\br} ) \right\rangle 
\label{eqn:lambda-eqn} \\
0 &=& \frac{3}{ 4 U} \ba_\tau
+ \frac{3 }{4 U} \operatorname{Im} \left\langle \tr ( \partial_\tau Z^\dagger_{\br} \bsigma Z^{\vphantom\dagger}_{\br} ) \right\rangle \nonumber  \\
&+& \frac{i}{2} \left\langle \psi^\dagger_{\br} \bsigma \psi^{\vphantom\dagger}_{\br} \right\rangle 
\label{eqn:atau-eqn}  \\
(\eta^{\br \br'})^\dagger &=& \left\langle Z^{\vphantom\dagger}_{\br'} Z^\dagger_{\br} \right\rangle \\
\eta^{\br \br'} &=& \frac{t_{\br \br'}}{|t_{\br \br'}|} \left\langle F^\dagger_{\br} F^{\vphantom\dagger}_{\br'} \right\rangle \text{.}
\end{eqnarray}

Following Ref.~\onlinecite{sslee05}, it is useful to make an analytic continuation in order to access a broader class of saddle points.  We consider
\begin{eqnarray}
\theta_{\br \br'} &\to& i \tilde{\theta}_{\br \br'} \\
\lambda_{\br} &\to& - i \tilde{\lambda}_{\br} \\
d^{\br \br'} &\to& i \tilde{d}^{\br \br'}_i  \\
\ba_\tau (\br) &\to& i \tilde{\ba}_\tau(\br) \text{,}
\end{eqnarray}
and therefore
\begin{eqnarray}
\eta^{\br \br'} &\to& |\chi^Z_{\br \br'}| U^{\br \br'}_Z \\
(\eta^{\br \br'})^\dagger &\to& |\chi^F_{\br \br'}|  
(U^{\br \br'}_F )^\dagger = |\chi^F_{\br \br'}| U^{\br' \br}_F \text{,}
\end{eqnarray}
where $|\chi^Z_{\br \br'}| = |\chi_{\br \br'}| e^{- \tilde{\theta}_{\br \br'}}$,
 $|\chi^F_{\br \br'}| = |\chi_{\br \br'}| e^{\tilde{\theta}_{\br \br'}}$, $U^{\br \br'}_Z = \exp\big[ i (c^{\br \br'} - \tilde{d}^{\br \br'}_i) \sigma^i \big]$, and $U_F^{\br' \br} = (U^{\br \br'}_F)^\dagger = \exp\big[ -i (c^{\br \br'}_i + \tilde{d}^{\br \br'}_i ) \sigma^i \big]$.   The saddle-point free energy is now real, and
the saddle-point equations become
\begin{eqnarray}
1 &=& \frac{1}{2} \left\langle \tr (Z^\dagger_{\br} Z^{\vphantom\dagger}_{\br} ) \right\rangle \label{eqn:lambda-eqn-2}  \\
0 &=& \frac{3 i}{4 U} \tilde{\ba}_\tau (\br)
+ \frac{3}{4 U} \operatorname{Im} \left\langle \tr ( \partial_\tau Z^\dagger_{\br} \bsigma Z^{\vphantom\dagger}_{\br} ) \right\rangle \nonumber  \\
&+& \frac{i}{2} \left\langle \psi^\dagger_{\br} \bsigma \psi^{\vphantom\dagger}_{\br} \right\rangle  \label{eqn:atau-eqn-2} \\
|\chi^F_{\br \br'}| U_F^{\br' \br} &=& \left\langle Z^{\vphantom\dagger}_{\br'} Z^\dagger_{\br} \right\rangle \\
|\chi^Z_{\br \br'}| U_Z^{\br \br'} &=& \frac{t_{\br \br'}}{| t_{\br \br'} |} \left\langle F^\dagger_{\br} F^{\vphantom\dagger}_{\br'} \right\rangle \text{.}
\end{eqnarray}

These expectation values, and also the mean-field ground state energy (see Appendix~\ref{app:mft-equiv}), are evaluated using the saddle-point action $\tilde{S} = \tilde{S}_0 + \tilde{S}_F + \tilde{S}_Z$, where
\begin{eqnarray}
\tilde{S}_0 &=&\int d\tau \Big\{ \sum_{(\br, \br')} |t_{\br \br'}| |\chi^F_{\br \br'}| |\chi^Z_{\br \br'}| \tr (U^{\br' \br}_F U^{\br \br'}_Z) \nonumber \\ &-& \sum_{\br} \tilde{\lambda}_{\br} \Big\}  \label{eqn:aux-spa}\\
\tilde{S}_F &=& \int d\tau \Big\{
\sum_{\br} \psi^\dagger_{\br} \partial_\tau \psi^{\vphantom\dagger}_{\br}
- \frac{1}{2} \sum_{\br} \psi^\dagger_{\br} \big[ \tilde{\ba}_{\tau}(\br) \cdot \bsigma  \big] \psi^{\vphantom\dagger}_{\br} \nonumber \\
&+&  \sum_{(\br , \br')} t_{\br \br'} |\chi^F_{\br \br'}| \big( \psi^\dagger_{\br} U^{\br \br'}_F \psi^{\vphantom\dagger}_{\br'} + \text{H.c.} \big) \Big\}  \\
\tilde{S}_Z &=& \int d\tau \Big\{ \sum_{\br} \zeta^\dagger_{\br} \Big[ \frac{3}{2 U} \big[ - \partial_\tau^2 + 
\big( \tilde{\ba}_\tau(\br) \cdot \bsigma \big) \partial_\tau \big]
+ \tilde{\lambda}_{\br} \Big] \zeta_{\br} \nonumber \\
&-& \sum_{( \br, \br')} |t_{\br \br'}| |\chi^Z_{\br \br'}| \big( \zeta^\dagger_{\br} U^{\br \br'}_Z \zeta_{\br'} + \text{c.c.} \big) \Big\} \text{.}
\end{eqnarray}

\subsection{Mean-field phase diagram and relation to $\rU1$ calculation}
\label{sec:mft-phasediag}

Now we focus specifically on the honeycomb Hubbard model at half filling, where
we are interested in describing spin-liquid states adjacent to the Mott transition to the semimetal.  We restrict our attention to an ansatz of the form
\begin{eqnarray}
\label{eqn:ansatz-1}
U_Z^{\br \br'} &=& \left( \begin{array}{cc}
e^{i \tilde{a}^Z_{\br \br'}} & 0 \\
0 & e^{-i \tilde{a}^Z_{\br \br'}}
\end{array} \right) \\
U_F^{\br \br'} &=& \left( \begin{array}{cc}
e^{i \tilde{a}^F_{\br \br'}} & 0 \\
0 & e^{-i \tilde{a}^F_{\br \br'}}
\end{array} \right) \\
\tilde{\ba}_\tau(\br) &=& 0 \text{.}
\label{eqn:ansatz-2}
\end{eqnarray}
$|\chi^F_{\br \br'}|$ and $|\chi^Z_{\br \br'}|$ are arbitrary.
While it might be interesting to determine the mean-field phase diagram considering a completely general ansatz, we have not done this.  First, we are interested in the possibility of a continuous transition from the semimetal to a spin liquid Mott insulator, and as we shall see the above ansatz allows us to describe the simplest scenario in which this can happen.  Second, the mean-field theory only gives very crude information about the energetics of various competing states, and is probably only useful as a rough guide to numerical simulations.  Therefore the value of a more complete mean-field study is rather dubious, at least until more numerical data are available.

For the ansatz considered, it turns out that the mean-field calculation is equivalent to that already carried out for the $\rU1$ slave rotor formulation.\cite{sslee05}  This equivalence is demonstrated in Appendix~\ref{app:mft-equiv}; the discussion there shows that the results from the $\rU1$ calculation can be used here with no change except a rescaling of the onsite interaction by $U \to (3/4) U$.
It should be stressed that, while the mean-field calculations in the $\rU1$ and $\rSU2$ formulations are formally equivalent (for this restricted class of ansatz), the interpretation of the resulting phase diagram in the two cases is different.
In particular, the spin liquid state we find here is described by Dirac fermions coupled to an $\rSU2$ gauge field, \emph{not} a $\rU1$ gauge field as was found in Ref.~\onlinecite{sslee05}.

Now we discuss the mean-field phase diagram, using the results of Ref.~\onlinecite{sslee05}.  We consider both nearest-neighbor ($t$) and third-neighbor ($t''$) hopping.  For $t'' = 0$, the $\rSU2$ ASL is the lowest-energy mean-field state for $1.26 < U / t < 1.30$.  For $U/t < 1.26$ the semimetal is the ground state; this can be determined by noticing that the gap to $Z$-boson excitations vanishes as $U/t$ is taken to $1.26$ from above.  For $U/t > 1.30$, a valence-bond solid is the ground state.  The third neighbor hopping helps stabilize the spin liquid; for $t'' = -0.4 t$, the window of $\rSU2$ ASL expands to
$1.7 < U/t < 1.9$.

In the case $t'' = 0$, the mean-field critical value $(U/t)_c = 1.26$ is significantly less than the value $(U/t)_c \approx 4-5$ determined by quantum Monte Carlo simulation and series expansion.\cite{paiva05}  Of course, it is not expected that the mean field theory will be quantitatively accurate, and it may not be qualitatively accurate.  However, it is reasonable to use the mean-field results as a rough guide to the energetics of the spin liquid state and some of its competitors, so it is a useful conclusion that adding third-neighbor hopping is likely to enhance the stability of the spin liquid.

We note that for $| t'' / t |$ large enough, the band structure changes \emph{qualitatively} from the $t'' = 0$ case.  This can be understood by considering the case $t =0$, $t'' \neq 0$, where the honeycomb lattice breaks into four larger honeycomb lattices, leading to \emph{eight} Dirac nodes at the Fermi level as opposed to two when $t \neq 0$, $t'' \neq 0$.  Straightforward calculation shows that the two-node band structure of $t'' = 0$ is confined to the region $-1 < t'' / t < 1/3$, with eight nodes present elsewhere.  The free-electron Hamiltonian with $t'' / t = -0.4$ lies safely in the two-node region, but this is not \emph{a priori} obvious for the mean-field spin liquid state obtained for the same parameters.  However, it turns out that, for these parameters, the ratio of spinon hopping matrix elements lies in the range $-0.4 < t''_s / t_s < 0$,\cite{sslee-pc} and the spin liquid state thus has only two nodes.  

Although we have not pursued this here, it would be interesting to investigate the possibility of spin liquid physics at \emph{larger} $|t'' / t|$.  One might be able to obtain a $\rSU2$ ASL with $N_f = 8$ Dirac fermions, which should be substantially more stable than the $N_f = 2$ case found here (at least as far as local stability to weak perturbations).  The semimetal state will also be different in this regime, with eight nodes instead of two.

\section{Low-energy effective theory}
\label{sec:effective-theory}

We shall write down a low-energy theory for the honeycomb Hubbard model describing the $\rSU2$ ASL, the semimetal and the Mott transition between them, as well as other nearby phases and critical points.  This effective theory contains the fermionic spinons and rotor bosons coupled to an $\rSU2$ gauge field.  Working on a space-time lattice, the action is written down in Sec.~\ref{sec:basic-effective-theory}, and the phase diagram is discussed.  Next, in Sec.~\ref{sec:spin-liquid-ft} the continuum quantum field theory for the ASL phase is obtained, and some of its physical properties are discussed.  The continuum theory of the Mott transition itself is constructed in Sec.~\ref{sec:mott-trans-ft}.  Finally, in Sec.~\ref{sec:u1sl} we consider the effect of breaking the pseudospin symmetry down to the ${\rm O}(2)$ subgroup generated by particle-hole symmetry and $\rU1$ charge rotations.  For sufficiently strong pseudospin-breaking of this type, we find a phase transition from the $\rSU2$ ASL to a $\rU1$ ASL, and recover the results of the $\rU1$ slave-rotor formulation.

\subsection{Gauge theory description of spin liquid and semimetal}
\label{sec:basic-effective-theory}

Here we construct the low-energy effective theory and use it to give a description of the spin liquid and semimetal phases.  In principle we could do this, starting in the spin liquid phase, by doing a detailed expansion about the mean-field saddle point.  However, because we expect that all fluctuations about this saddle point will be massive, except for those captured as an $\rSU2$ gauge boson, we can simply write down an $\rSU2$ gauge theory.  It is convenient to discretize imaginary time and work on a space-time lattice, going to a continuum formulation later on.

We consider a Euclidean space-time of stacked honeycomb lattices, with lattice sites labeled by
by $x = (\br, \tau) = \br + \tau \bz$, where $\tau = \epsilon n$ is a discrete imaginary time ($n$ is an integer and $\epsilon$ is the time step), $\br$ labels the sites of the honeycomb lattice, and $\bz$ is a unit vector pointing along the imaginary time direction.  On each lattice point we have an $\rSU2$ matrix $Z_x$ and the $2 \times 2$ matrix of fermions
\begin{equation}
F_x = \left( \begin{array}{cc}
f_{\br \uparrow}(\tau) & \epsilon_{\br} \bar{f}_{\br \downarrow}(\tau) \\
f_{\br \downarrow}(\tau) & -\epsilon_{\br} \bar{f}_{\br \uparrow}(\tau)
\end{array} \right) \text{.}
\end{equation}
On nearest-neighbor bonds within each honeycomb layer (``spacelike'' bonds), as well as on all ``timelike'' bonds connecting two adjacent layers,  we place the $\rSU2$ gauge field $U_{x x'}$.  The action takes the form $S^{{\rm eff}} = S^{{\rm eff}}_F + S^{{\rm eff}}_Z + S^{{\rm eff}}_g$, and the partition function is
\begin{equation}
Z^{{\rm eff}} = \int {\cal D}Z {\cal D} \bar{f} {\cal D}f {\cal D} U e^{-S^{{\rm eff}}} \text{.}
\end{equation}
Here,
\begin{eqnarray}
S^{{\rm eff}}_F  &=&  t_s^\tau \sum_x \tr \big( F_{x + \epsilon \bz} U_{x + \epsilon \bz, x} F^\dagger_{x} \big) \nonumber \\
&+& t^{\br}_s \sum_\tau \sum_{\langle \br \br' \rangle} \tr \big( F_{x} U_{x x'} F^\dagger_{x} \big) \label{eqn:seff-f-eqn} \text{,}
\end{eqnarray}
where it should be noted that the first term defines an orientation on the timelike bonds; that is, the overall sign is reversed if $x$ and $x+\epsilon \bz$ are interchanged.  It is worth noting that this does \emph{not} break time-reversal symmetry, as time-reversal maps the \emph{imaginary} time $\tau$ to itself, not to $- \tau$.
The second term, consistent with the spatial reflection symmetries of the honeycomb lattice, does not introduce an orientation for the spacelike bonds. Next we have
\begin{eqnarray}
S^{{\rm eff}}_Z &=& -t_c^{\tau} \sum_{x} \tr \big( Z^\dagger_{x + \epsilon \bz} U_{x + \epsilon \bz, x} Z_x \big) \nonumber \\
&-& t_c^{\br} \sum_\tau \sum_{\langle \br \br' \rangle} \tr \big( Z^\dagger_x U_{x x'} Z_x' \big)
\label{eqn:seff-z-eqn} \text{,}
\end{eqnarray}
and
\begin{equation}
S^{{\rm eff}}_g = -K \sum_p \Big( \tr \big( U_{x x'} U_{x' x''} \dots \big)  + \text{c.c.} \Big) \label{eqn:seff-g-eqn} \text{,}
\end{equation}
where the sum is over two types of lattice plaquettes: hexagonal plaquettes lying within the honeycomb layers, and 4-sided plaquettes containing two timelike and two spacelike bonds.   The trace is over a product of $U_{x x'}$, which is taken moving along the plaquette edge in an arbitrary direction.

Upon taking the time-continuum limit $\epsilon \to 0$, we recover the correct dynamics for the fermions and rotor bosons.  We focus on the terms involving timelike bonds in $S^{{\rm eff}}_F$ and $S^{{\rm eff}}_Z$.  Letting $t_s^\tau = - 1/2$, $t_c^\tau = 3 / ( 2 U \epsilon)$ and 
\begin{equation}
U_{x + \epsilon \bz, x} = \exp \Big( - \epsilon \frac{ i \ba_\tau \cdot \bsigma}{2} \Big) \text{,}
\end{equation}
and taking the limit $\epsilon \to 0$, the time-derivative terms
\begin{equation}
\frac{1}{2} \int d\tau \sum_{\br} \tr \Big[ F_{\br} \big( \partial_\tau + \frac{i \ba_\tau \cdot \bsigma}{2} \big) F^\dagger_{\br} \Big]
\end{equation}
and
\begin{equation}
\frac{3}{4U} \int d\tau \sum_{\br} \tr \Big[ Z^\dagger_{\br} \big(\overleftarrow{\partial}_\tau - \frac{i \ba_\tau \cdot \bsigma}{2} \big) \big( \partial_\tau + \frac{i \ba_\tau \cdot \bsigma}{2} \big) Z_{\br} \Big]
\end{equation}
result.  These are identical to those derived microscopically in Sec.~\ref{sec:su2-formulation}.  

\begin{figure}
\includegraphics[width=3in]{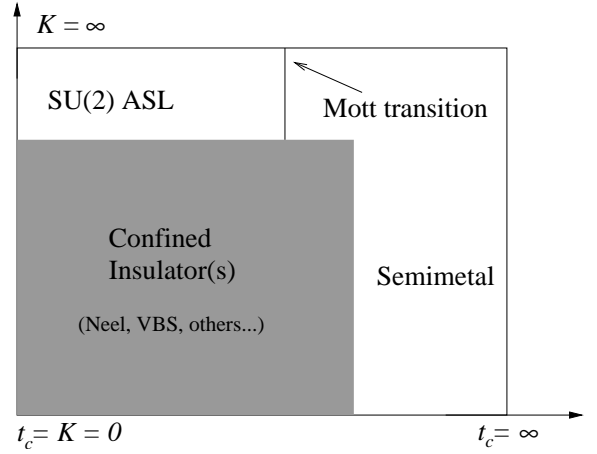}
\caption{Schematic phase diagram of the effective low-energy theory $S^{{\rm eff}}$ given in Eqs.~(\ref{eqn:seff-f-eqn}), (\ref{eqn:seff-z-eqn}) and~(\ref{eqn:seff-g-eqn}).  The fermion hopping $t_s$ has been set to unity by rescaling the fermion fields.}
\label{fig:seff-phasediag}
\end{figure}

We now discuss the phase diagram of the effective action.
We shall find it convenient to work with discrete imaginary time (\emph{i.e.} finite $\epsilon$).  For simplicity, we shall set $t_c^\tau = t_c^{\br} = t_c$ and $- t_s^\tau = t_s^{\br} = t_s$,  and we may set $t_s =1$ without further loss of generality, by rescaling the fermion fields.  The phase diagram may now be drawn as a function of the dimensionless parameters $t_c$ and $K$; this is shown schematically in Fig.~\ref{fig:seff-phasediag}.

We are most interested in understanding what happens when $K$ is large.  In this limit the gauge fluctuations are suppressed, and the spinons and rotor bosons become ``good variables'' with which to understand the physics.  Furthermore, in this limit we will be able to access the $\rSU2$ ASL, the semimetal, and a continuous quantum phase transition between them.

We first consider the limit $K \gg 1$ and $t_c \ll 1$.  Here, the rotor bosons will be gapped, and we can safely integrate them out to obtain an action involving only the fermions and the gauge field.
At large $K$, the term $S^{{\rm eff}}_g$ favors gauge field configurations that are gauge-equivalent to $U_{x x'} = 1$.  If we set $K$ to infinity we can drop the gauge field altogether in $S^{{\rm eff}}_F$, and the spinons behave as free Dirac fermions.  For $K$ large but finite, these Dirac fermions become coupled to a fluctuating $\rSU2$ gauge field -- we are therefore describing an $\rSU2$ algebraic spin liquid.

The simplest possibility is that the $\rSU2$ ASL is a stable phase.  While this assumption can be justified in the limit of a large number of fermion flavors,\cite{ran06} we are not guaranteed that it holds for the Hubbard model.  As we shall discuss in Sec.~\ref{sec:discussion}, even if it is not stable, the $\rSU2$ ASL may still have important implications for the metal-insulator transition.  For the moment we simply assume that the $\rSU2$ ASL is a stable phase.

Next, we consider the limit $K, t_c \gg 1$.  The rotor bosons are condensed, resulting in a Higgs phase for the $\rSU2$ gauge field.  To understand the nature of this phase, we couple an external electromagnetic vector potential $A_{x x'}$ to the rotor bosons.  Their part of the action then takes the form
\begin{equation}
\label{eqn:seff-z-with-emfield}
S^{{\rm eff}}_Z = - t_c \sum_{\langle x x' \rangle} \tr \big(e^{ i A_{x x'} \tau^3} Z^\dagger_x U_{x x'} Z_{x'}   \big) \text{.}
\end{equation}
This form can be understood by noting that $A_{x x'}$ is minimally coupled to the $\tau^3$-component of the pseudospin, which is simply the electromagnetic charge.  Now we make the $\rSU2$ gauge transformation (to ``unitary gauge'')
\begin{eqnarray}
U_{x x'} &\to& Z_x U_{x x'} Z^\dagger_{x'} \\
F_x &\to& F_x Z^\dagger_x \text{.}
\end{eqnarray}
The various terms in $S^{{\rm eff}}$ now take the form
\begin{eqnarray}
S^{{\rm eff}}_Z &=& - t_c \sum_{\langle x x' \rangle} \tr \big( e^{i A_{x x'} \tau^3} U_{x x'} \big) \\
S^{{\rm eff}}_F &=& \sum_{\langle x x' \rangle} \tr \big( F_x U_{x x'} F^\dagger_{x'} \big) \text{,}
\end{eqnarray}
and the form of $S^{{\rm eff}}_g$ is unchanged. When $t_c$ is large, $U_{x x'}$ has only small fluctuations about $e^{- i A_{x x'} \tau^3}$, so we write
\begin{equation}
U_{x x'} = \exp \Big(- i A_{x x'} \tau^3 \Big) \exp \Big(  \frac{i \tilde{\ba}_{x x'} \cdot \btau}{2} \Big) \text{,}
\end{equation}
where $\tilde{\ba}_{x x'}$ may be treated as a small quantity.  We can then expand $S^{{\rm eff}}_Z$ for small $\tilde{\ba}_{x x'}$, and find (at leading non-vanishing order) a quadratic ``mass'' term for the fluctuations of the gauge field.  Therefore the gauge field is massive in this phase, and can be integrated out.  The fermion part of the action becomes
\begin{eqnarray}
S^{{\rm eff}}_F &=& \sum_{\langle x x' \rangle} \tr \big( F_x e^{- i A_{x x'} \tau^3} e^{i \tilde{\ba}_{x x'}\cdot\btau/2} F_x' \big) \\
&\approx& \sum_{\langle x x' \rangle} \tr \big( F_x e^{-i A_{x x'} \tau^3} F_x' \big) \text{.}
\end{eqnarray}
In the second line, the coupling to the massive gauge field has been dropped, as it will only generate irrelevant four-fermion interactions.  Therefore the low-energy theory consists simply of free Dirac fermions minimally coupled to the electromagnetic field.  In fact, going to temporal gauge and taking the time-continuum limit we can rewrite this as
\begin{equation}
S^{{\rm eff}}_F \to \int d\tau \Big[ \sum_{\br} \bar{f}_{\br \alpha} \partial_\tau f_{\br \alpha}
- t_s \sum_{\langle \br \br' \rangle} \big( \bar{f}_{\br \alpha} e^{i A_{\br \br'}} f_{\br' \alpha} + \text{H.c.} \big) \Big] \text{,}
\end{equation}
which shows that these fermions are just electrons, and we are describing the semimetal.

It should be noted that, even though we have condensed the pseudospin-carrying field $Z_x$, 
pseudospin is not broken in the Higgs phase, as is correct for the semimetal.  To see this, we work for simplicity in the $K = \infty$ limit,\footnote{The finite $K$ case can be understood in essentially in the same manner.  The only change is that the gauge fixing becomes more complicated, but can be handled following the procedure of Ref.~\onlinecite{creutz77}}
where $U_{x x'}$ is pure gauge, and we may choose $U_{x x'} = 1$.  
Then, in the Higgs phase,
\begin{equation}
\langle Z_x \rangle = 
\Big\langle \left( \begin{array}{cc}
z_{x 1} & z_{x 2} \\
-z^*_{x 2} & z^*_{x 1}
\end{array} \right) \Big\rangle
\equiv \bar{Z}_x =
\left( \begin{array}{cc}
\bar{z}_{1} & \bar{z}_{2} \\
-\bar{z}^*_{2} & \bar{z}^*_{1}
\end{array} \right) \text{.}
\end{equation}
In this gauge, the ``condensate'' $\langle Z_x \rangle$ is independent of position $x$.
By assumption, at least one of $\bar{z}_1, \bar{z}_2$ is nonzero, so $ \bar{Z}_x $ can be written as a real number times an $\rSU2$ matrix.  If instead we work in the gauge where $U_{x x'} = \Lambda_x \Lambda^\dagger_{x'}$, where $\Lambda_x$ are $\rSU2$ matrices, then  
$\langle Z_x \rangle = \Lambda_x \bar{Z}_x$.  Therefore, any two condensates can be rotated into one another by an appropriate $\rSU2$ gauge transformation.

Working in any fixed gauge, it might appear that the pseudospin symmetry has been broken, because $\bar{Z}_x$ will transform nontrivially under pseudospin rotation.  Consider a condensate specified by $\bar{Z}^1_x$ that can be transformed into $\bar{Z}^2_x$ via a global pseudospin rotation.  We can also transform $\bar{Z}^1_x$ into $\bar{Z}^2_x$ by making an $\rSU2$ gauge transformation. Now, configurations that differ only by a gauge transformation should be thought of as different \emph{labels} for the \emph{same} physical state;\cite{wen02} therefore, $\bar{Z}^1_x$ and 
$\bar{Z}^2_x$ both describe the same state, and the apparent pseudospin breaking is a gauge artifact.

Next, the Mott transition itself can be understood starting from the spin liquid phase.  Barring a first-order transition, as $t_c$ is increased, the gap to rotor boson excitations will go to zero at a critical value of $t_c$.  The bosons condense on the other side of the critical point, and the semimetal is obtained.

We also briefly discuss the other regions of the phase diagram, although these are not our main focus.  For $t_c \gg 1$ and arbitrary $K$, we still expect to be in the semimetal phase -- this is clearest if one works in unitary gauge as above.  For $K, t_c \ll 1$ the gauge field fluctuates strongly and the physics is less clear, especially given the coupling to the fermions.  This region of the phase diagram is expected to be occupied primarily by insulating states where the gauge field is in a confining phase.  The most likely such state, which is expected to occupy at least some part of the phase diagram, is the N\'{e}el antiferromagnet.  Valence-bond solids and other states are also possible.

Finally, we are now in a position to comment on the analysis of Refs.~\onlinecite{kskim06a, kskim06b}, where it was claimed that the $\rSU2$ slave rotor formulation can lead to a description of a non-Fermi liquid metallic state.  In terms appropriate to the present discussion and notation, that analysis identified the non-Fermi liquid as a state where, in some fixed gauge, $\langle z_1 \rangle \neq 0$ and $\langle z_2 \rangle = 0$.  The Fermi liquid was identified as a state where both $\langle z_1 \rangle$ and $\langle z_2 \rangle$ are nonzero.  However, as per the discussion above, these two patterns of boson condensation can be rotated into one another by an $\rSU2$ gauge transformation and thus do not describe distinct phases.  This is true even away from half-filling where the pseudospin symmetry is broken, which is the situation considered in Refs.~\onlinecite{kskim06a,kskim06b}.  In general, we feel it is probably not possible to describe a non-Fermi liquid conducting state in the $\rSU2$ slave-rotor formulation.

\subsection{Field theory for the spin liquid}
\label{sec:spin-liquid-ft}

We now construct a continuum quantum field theory for the $\rSU2$ ASL; many results stated in this section are discussed in more detail in Appendix~\ref{app:field-theory}.  We begin with the continuum description of the spin liquid within mean-field theory.  On the lattice the mean-field Hamiltonian is 
\begin{equation}
{\cal H}_{{\rm MFT}} = 
t_s \sum_{\langle \br \br' \rangle} \big( \psi^\dagger_{\br} \psi^{\vphantom\dagger}_{\br'} + \text{H.c.} \big)
\text{,}
\end{equation}
where $\psi_{\br}$ is defined in Eq.~(\ref{eqn:psi-defn}).
We have included nearest-neighbor hopping only; as long as one allows arbitrary perturbations to the continuum effective action, this is sufficient to derive the low-energy theory.
As discussed in Sec.~\ref{sec:intro} and Fig.~\ref{fig:honeycomb}, the $\rSU2$ spinors $\psi_{\br}$ are also labeled as $\psi_{\bR i}$, where $\bR$ labels the unit cells of the honeycomb lattice and $i = 1, 2$ labels the two-site basis.  The Fourier transform is then defined as
\begin{equation}
\psi_{\bk i} = \frac{1}{\sqrt{N_c}} \sum_{\bR} e^{-i \bk \cdot \bR} \psi_{\bR i} \text{,}
\end{equation}
where $N_c$ is the number of unit cells in the lattice.

The Hamiltonian ${\cal H}_{{\rm MFT}}$ has gapless Dirac points at the Fermi level (fixed at zero energy by spinon particle-hole symmetry, which is part of the $\rSU2$ gauge group).  These nodal points lie 
at $\pm \bQ$ in the Brillouin zone, where $\bQ = (4 \pi / 3) \bx$.  Working in momentum space, we define the continuum fields
\begin{eqnarray}
\varphi_1(\bq) &\sim& \left( \begin{array}{c}
\psi_{\bQ + \bq, 1} \\
- \psi_{\bQ + \bq, 2} \end{array} \right) \\
\varphi_2(\bq) &\sim& \left( \begin{array}{c}
\psi_{-\bQ + \bq, 2} \\
\psi_{-\bQ + \bq, 1}
\end{array} \right) \text{,}
\end{eqnarray}
where $|\bq| \ll \Lambda$, and $\Lambda$ is a cutoff much smaller than the zone size.  We can also Fourier transform to obtain real-space continuum fields
\begin{equation}
\varphi_a(\br) = \int \frac{d^2 \bq}{(2 \pi)^2} e^{i \bq \cdot \br} \varphi_a (\bq) \text{,}
\end{equation}
where the momentum cutoff $\Lambda$ is implicit.

The field $\varphi_a$ (where $a = 1,2$) is actually a four-component object, and it is convenient to define two kinds of Pauli matrices acting on it by defining the following $4 \times 4$ matrices in block form:
\begin{eqnarray}
\mu^i &\equiv& \left( \begin{array}{cc}
\mu^i & 0 \\
0 & \mu^i \end{array} \right) \\
\tau^i &\equiv& \left( \begin{array}{cc}
\tau^i_{1 1} & \tau^i_{1 2} \\
\tau^i_{2 1} & \tau^i_{2 2}
\end{array} \right) \\
\tau^i \mu^j = \tau^i \otimes \mu^j &\equiv& 
\left( \begin{array}{cc}
\tau^i_{1 1} \mu^j & \tau^i_{1 2} \mu^j \\
\tau^i_{2 1} \mu^j & \tau^i_{2 2} \mu^j
\end{array} \right) \text{.}
\end{eqnarray}
We say that the $\mu^i$ Pauli matrices act in the $\rSU2$ gauge space, and the $\tau^i$ Pauli matrices act in the Dirac spinor or Lorentz space.

The continuum mean-field Lagrangian density (in imaginary-time) is
\begin{equation}
{\cal L}_0 = \bar{\varphi}_a \big[ - i \gamma_{\mu} \partial_\mu \big] \varphi_a \text{,}
\end{equation}
where $\mu = 0,1,2$, we have defined
\begin{equation}
\bar{\varphi}_a = i \varphi^\dagger_a \tau^3 \text{,}
\end{equation}
and
\begin{equation}
\gamma_\mu = (\tau^3, \tau^2, -\tau^1) \text{.}
\end{equation}
This is nothing but the Lagrangian for massless Dirac fermions in $2+1$ dimensions.  The coupling to the fluctuating $\rSU2$ gauge field is introduced by replacing $\partial_\mu$ with the covariant derivative
\begin{equation}
D_{\mu} \equiv \partial_\mu + \frac{ i a^i_{\mu} \mu^i}{2} \text{.}
\end{equation}
We also define the field-strength tensor for the gauge field as 
\begin{equation}
f^i_{\mu \nu} = \partial_\mu a^i_{\nu} - \partial_\nu a^i_{\mu} + \epsilon^{i j k} a^j_{\mu} a^k_{\nu} \text{.}
\end{equation}
The effective Lagrangian for the $\rSU2$ ASL is then
\begin{equation}
\label{eqn:asl-effective-L}
{\cal L}^{{\rm eff}}_{{\rm ASL}} = \bar{\varphi}_a \big[ -i \gamma_\mu D_\mu \big] \varphi_a
+ \frac{1}{4 g^2} f^i_{\mu \nu} f^i_{\mu \nu} \text{.}
\end{equation}
The gauge coupling constant $g^2$ has dimensions of mass. Other terms consistent with the symmetries of the underlying model are added as perturbations, and local stability against such perturbations can be considered.

As it stands this field theory is strongly coupled and cannot be analyzed reliably -- a straightforward perturbation theory leads to severe infrared divergences.  However a controlled analysis is possible in a large-$N_f$ expansion,\cite{appelquist90, ran06} where instead of $a = 1,2$,  one allows $a = 1, \dots, N_f$.  (In our notation, the physical case corresponds to $N_f = 2$.)  The perturbation theory is reorganized into an expansion in powers of $1/N_f$, whereupon the infrared divergences are cured and systematic calculation becomes possible.  As in the case of the $\rU1$ ASL,\cite{appelquist86,rantner01,rantner02} one arrives at the conclusion that the $\rSU2$ ASL is an interacting critical state, with nontrivial critical exponents that vary as a function of $1/N_f$.  Also as in the $\rU1$ case,\cite{hermele04} this state is a stable phase in the large-$N_f$ limit; the only potentially dangerous perturbations are fermion mass terms or ``velocity anisotropy'' terms (\emph{i.e.} fermion bilinears with one derivative), and on the honeycomb lattice it is easy to show that all such terms are forbidden by symmetry.

\begin{figure}
\includegraphics[width=3in]{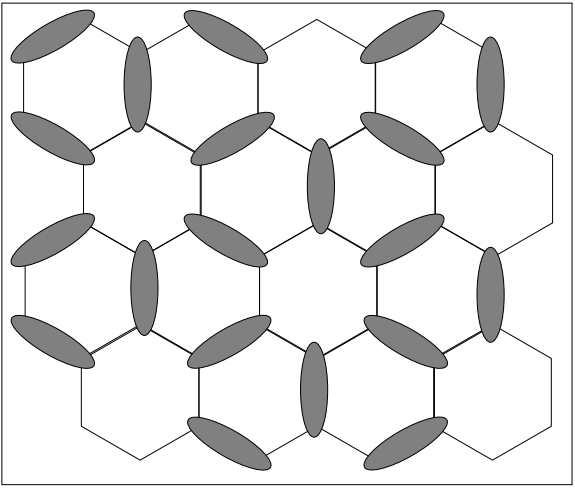}
\caption{Depiction of a valence-bond solid state on the honeycomb lattice. The shaded regions represent spin-singlet valence bonds.  This VBS pattern corresponds to one of the fluctuating competing orders present in the $\rSU2$ ASL; the other such competing order is simply N\'{e}el antiferromagnetism.  
Both these observables have slowly decaying power-law correlations characterized by the \emph{same} critical exponent; this phenomenon should be a useful test for the $\rSU2$ ASL in future numerical studies.}
\label{fig:hcvbs}
\end{figure}

Very recently, in the context of an $\rSU2$ ASL in the square lattice Heisenberg model (the $\pi$-flux state\cite{affleck88, marston89}), Ran and Wen have shown that the $\rSU2$ ASL has an emergent ${\rm Sp}(2 N_f)$ symmetry.\cite{ran06}  This symmetry is present exactly in Eq.~(\ref{eqn:asl-effective-L}), but it is hidden and some work is needed to expose it (see Ref.~\onlinecite{ran06} and Appendix~\ref{app:field-theory} for details).  The ${\rm Sp}(2 N_f)$ plays a role analogous to the ${\rm SU}(N_f)$ emergent symmetry recently discussed in the context of $\rU1$ algebraic spin liquids;\cite{hermele05} in particular, it also leads to a unification of seemingly unrelated ``competing orders.''  These are a set of observable quantities that have slowly decaying power-law correlations, all with the same critical exponent.  Ref.~\onlinecite{ran06} found that a set of five fermion bilinears are unified together into an ${\rm SO}(5)$ vector representation  of the ${\rm Sp}(4)$ symmetry (recall that ${\rm Sp}(4)$ and ${\rm SO}(5)$ have the same Lie algebra), and that coupling to the $\rSU2$ gauge field results in slowly decaying correlations for this ${\rm SO}(5)$ vector.  For the square lattice Heisenberg model, they found that three components of the vector correspond to the N\'{e}el order parameter, and the other two to the order parameter for columnar and box valence-bond solid (VBS) states.\cite{read90}  

The result for the honeycomb lattice is very similar; the ${\rm SO}(5)$ vector again contains the N\'{e}el order parameter $\boldsymbol{n}$ and the two-component order parameter for a VBS state, $\boldsymbol{v} = (v_1, v_2)$.  In Appendix~\ref{app:field-theory} these fields are defined as gauge-invariant bilinears of the $\varphi_a$ fermion fields.  The VBS order parameter corresponds to the pattern of dimerization shown in Fig.~\ref{fig:hcvbs}.  Using the result of Ref.~\onlinecite{ran06}, to leading order in the $1/N_f$ expansion the scaling dimension of these operators is
\begin{equation}
\Delta \equiv \operatorname{dim} [ \boldsymbol{n} ] = \operatorname{dim} [ v_i ] = 2 - \frac{32}{\pi^2 N_f}
\text{.}
\end{equation}
This means that N\'{e}el correlations, and the VBS correlations pictured in Fig.~\ref{fig:hcvbs}, fall off with the same power-law decay at long distances.  For instance,
\begin{equation}
\langle \bS_{\bR 1} \cdot \bS_{\boldsymbol{0} 1} \rangle \sim \frac{1}{|\bR|^{2 \Delta}} \text{,}
\end{equation}
and
\begin{equation}
\big\langle (\bS_{\bR 1} \cdot \bS_{\bR 2}) (\bS_{\boldsymbol{0} 1} \cdot \bS_{\boldsymbol{0} 2}) \big\rangle - \big( \langle \bS_{\boldsymbol{0} 1} \cdot \bS_{\boldsymbol{0} 2} \rangle \big)^2
 \sim \frac{\cos(4 \pi R_x / 3)}{|\bR|^{2 \Delta}} \label{eqn:ex-vbs-correlator} \text{.}
\end{equation}
It should be noted that the magnitude of the $1/N_f$ correction to $\Delta$ is rather large, so the above correlators may decay quite slowly, making them easy to observe in numerical simulations.
Furthermore, even if the system sizes available are not sufficient to determine whether the critical exponents for N\'{e}el and VBS correlations are the same, the simultaneous presence of slowly decaying correlations in both observables would already be a sign of the $\rSU2$ ASL.

It may be useful to mention that VBS correlations can be measured in two different ways.  The first is, of course, to look at a dimer-dimer correlation function as in Eq.~(\ref{eqn:ex-vbs-correlator}).  However, because there is no difference at the level of symmetry between the two quantities, one can just as well look as correlations of the bond kinetic energy density.  This would be simpler in determinantal quantum Monte Carlo simulations, because the relevant correlation function involves only four electron operators, as opposed to eight in the dimer-dimer correlator. The kinetic energy on the bond $(\br, \br')$ is
\begin{equation}
\hat{K}_{\br \br'} = - ( c^\dagger_{\br \alpha} c^{\vphantom\dagger}_{\br' \alpha} + \text{H.c.} ) \text{,}
\end{equation}
and an example of its long-distance correlation function in the $\rSU2$ ASL is
\begin{equation}
\langle \hat{K}_{(\bR 1, \bR 2)} \hat{K}_{(\boldsymbol{0} 1, \boldsymbol{0} 2)} \rangle
- \big( \langle \hat{K}_{(\boldsymbol{0} 1, \boldsymbol{0} 2)} \rangle \big)^2
\sim 
\frac{\cos(4 \pi R_x / 3)}{|\bR|^{2 \Delta}} \text{,}
\end{equation}
with the same critical exponent $\Delta$ as above.
 
\subsection{Field theory for the Mott transition}
\label{sec:mott-trans-ft}

To construct the continuum field theory for the Mott transition, we start with that discussed above for the spin liquid state, and add in the rotor bosons.  To understand the form of the bosonic sector, we again start at the mean-field level.  On the lattice, the bosons obey the mean-field action
\begin{eqnarray}
\label{eqn:sz-mft-action}
S^Z_{{\rm MFT}} &=& - \sum_{\tau} \sum_{\br} \Big( \tr (Z^\dagger_{x + \epsilon \bz} Z_{x}) + \text{c.c.} \Big) \nonumber \\
&-& \sum_{\tau} \sum_{\langle \br \br' \rangle} \Big( \tr(Z^\dagger_{(\br, \tau)} Z_{(\br', \tau)}
+ \text{c.c.} \Big) \nonumber \\
&+& \frac{r + r_{c0}}{2} \sum_{\tau, \br} \tr (Z^\dagger_x Z_x ) \text{,}
\end{eqnarray}
where for simplicity the coefficients of the first two terms have been set to unity, and
\begin{equation}
Z_x = \left( \begin{array}{cc}
z_{x 1} & z_{x 2} \\
- z^*_{x 2} & z^*_{x 1}
\end{array} \right) \text{,}
\end{equation}
where $z_{x 1}, z_{x 2}$ are \emph{arbitrary} complex numbers.  The constant $r_{c0}$ is chosen so that the lowest energy mode of the boson field just becomes gapless as $r$ is decreased to zero.
It is a straightforward task to diagonalize the quadratic form of Eq.~(\ref{eqn:sz-mft-action}), and one finds that for $r > 0$ there is a single low-energy mode at $\bk = 0$ in the Brillouin zone.  The low-energy fluctuations can be described by a continuum $2 \times 2$ matrix field $\Theta$, which can be related to the lattice fields by
\begin{equation}
\Theta(x) \sim Z_{[\tau, (\bR ,1)]} + Z_{[\tau, (\bR, 2)]}  \text{.}
\end{equation}
Here, $(\bR ,1)$ and $(\bR, 2)$ label the two honeycomb sites in the unit cell at $\bR$.   
We write 
\begin{equation}
\Theta = \left(\begin{array}{cc}
\theta_1 & \theta_2 \\
-\theta_2^* & \theta_1^* 
\end{array}\right) \text{,}
\end{equation}
where $\theta_1, \theta_2$ are arbitrary complex numbers.

The mean-field Lagrangian density for $\Theta$ is
\begin{equation}
{\cal L}^{\Theta}_0 = \frac{1}{2} \tr \big( \Theta^\dagger \overleftarrow{\partial}_\mu \partial_\mu \Theta \big) + \frac{r}{2} \tr \big( \Theta^\dagger \Theta \big) \text{,}
\end{equation}
where, in the first term, the first partial derivative acts to the left.  We have set the velocity of the $\Theta$ boson to unity -- we come back to this point shortly.   It should be noted that no linear time derivative term is present.  Such a term is forbidden by particle-hole symmetry (together with invariance under global $\rSU2$ gauge transformations and $\rU1$ charge rotations).
This has the important consequence that the bosonic sector obeys relativistic dynamics at low energy.

The coupling to the gauge field is included by introducing covariant derivatives, and the kinetic energy term for $\Theta$ becomes
\begin{equation}
{\cal L}^{\Theta}_{{\rm kin}} = \frac{1}{2} \tr \Big( \Theta^\dagger \big( \overleftarrow{\partial}_{\mu} - \frac{i a^i_\mu \sigma^i}{2} \big) \big( \partial_\mu + \frac{i a^i_{\mu} \sigma^i}{2}\big) \Theta \Big) \text{.}
\end{equation}
The full effective action describing the Mott transition is then
\begin{equation}
{\cal L}^{{\rm eff}}_{{\rm Mott}} = {\cal L}^{{\rm eff}}_{{\rm ASL}} + {\cal L}^{\Theta}_{{\rm kin}}
+ \frac{r}{2} \tr \big( \Theta^\dagger \Theta \big) + \frac{\lambda}{4} \Big( \tr \big(\Theta^\dagger \Theta \big) \Big)^2 + \cdots \text{,}
\end{equation}
where we have also included a quartic term for the $\Theta$ boson field.  This can be written in a more conventional form if we define the two-component field
\begin{equation}
\vartheta = \left( \begin{array}{c}
\theta_1 \\
- \theta^*_2 
\end{array}\right) \text{.}
\end{equation}
The bosonic part of the Lagrangian then becomes
\begin{eqnarray}
{\cal L}^{\Theta}_{{\rm kin}}
&+& \frac{r}{2} \tr \big( \Theta^\dagger \Theta \big) + \frac{\lambda}{4} \Big( \tr \big(\Theta^\dagger \Theta \big) \Big)^2  \nonumber \\
&=& \Big|  \big( \partial_\mu + \frac{i a^i_\mu \sigma^i}{2} \big) \vartheta \Big|^2 +
r \vartheta^\dagger \vartheta + \lambda ( \vartheta^\dagger \vartheta )^2
\end{eqnarray}
So the full field theory is simply $N_f = 2$ Dirac fermions and $N_b = 1$ relativistic bosons, both coupled to an $\rSU2$ gauge field.  All the matter fields transform in the fundamental representation of the gauge group.

The Mott transition from the spin liquid will occur when $r$ is tuned through a critical value $r_c$ and the mass for $\Theta$ vanishes.  On the conducting side of the transition, $\Theta$ condenses and gives the gauge boson a mass via the Higgs mechanism.  As discussed above in Sec.~\ref{sec:basic-effective-theory}, this results in a description of the semimetal.  The critical properties of the Mott transition can be analyzed in at least two controlled limits:  one can make the number of both fermion ($N_f$) \emph{and} boson ($N_b$) fields large and perform a simultaneous expansion in $1/N_f$ and $1/N_b$, or one can carry out an expansion in $4 - \epsilon$ space-time dimensions.  As our primary focus is on signatures of the spin liquid phase itself, we have left such calculations for future work.  If evidence for spin liquid physics of the type we propose here is indeed found in the honeycomb Hubbard model, it would be  important to work out the properties of the critical point in more detail.

A word about Lorentz invariance is now in order.  There are three bare velocities in the effective Lagrangian -- that of the fermions, the $\Theta$ boson and the gauge boson -- and in order to have Lorentz invariance they must all be equal.  The bare gauge boson velocity is actually not important, because the long-distance, low-energy dynamics for the gauge bosons are generated entirely by the matter fields.  One of the remaining velocities can be set to unity by an appropriate choice of units, but the ratio of the fermion and boson velocities is a dimensionless parameter that cannot be removed by scaling, and in principle may be important at the Mott critical point.  Supposing the difference in velocities is small, it may be treated as a perturbation starting from the Lorentz-invariant critical point, which could be irrelevant, marginal or relevant in the RG sense.  While this can be resolved by RG calculations of the type described above, the answer is not immediately obvious, because, in both the $N_f, N_b \to \infty$ and $\epsilon \to 0$ limits, the difference in velocities is marginal, and its fate will be determined by perturbative corrections in the appropriate small parameters.  The most likely outcome, which occurs in other cases,\cite{vafek02, saremi06} is that the difference in velocities will be irrelevant and the system will flow to the Lorentz-invariant critical point.  This is an important issue; if the difference in velocities is \emph{relevant}, then even in the large-$N$ or small-$\epsilon$ limit the critical point will have two relevant directions and will be unstable.  While we feel this is a very unlikely scenario, this issue will need to be addressed in any future study of the critical properties.

\subsection{Pseudospin breaking and $\rU1$ ASL}
\label{sec:u1sl}

The $\rSU2$ slave-rotor formulation has allowed us to understand the Mott transition out of the semimetal into the $\rSU2$ ASL, so far assuming the presence of pseudospin symmetry.  On the other hand, the $\rU1$ slave-rotor formulation finds instead a Mott transition into a $\rU1$ ASL.\cite{sslee05}  While the $\rU1$ formulation does not respect the pseudospin symmetry, we might expect to recover its results in the presence of perturbations that break pseudospin down to the ${\rm O}(2)$ subgroup generated by $\rU1$ charge rotations and particle-hole symmetry.  It is interesting to explore this possibility because the $\rU1$ ASL is likely to be more stable than its $\rSU2$ counterpart, and also because it possesses a different pattern of competing orders with slowly decaying power-law correlations,\cite{hermele05} and can thus be easily distinguished from the $\rSU2$ ASL.    Furthermore,
this situation should be accessible to quantum Monte Carlo simulation without a sign problem, for example by adding the nearest-neighbor 
interaction ${\cal H}_V$ of Eq.~(\ref{eqn:hv-term}).  

Here, we consider the effect of adding a pseudospin-breaking term to the low-energy theory of Sec.~\ref{sec:basic-effective-theory}.  The term we choose to add is not simply ${\cal H}_V$ written in slave-rotor variables; however, its symmetry transformation properties are identical to those of ${\cal H}_V$, and it will be generated upon integrating out some high energy degrees of freedom.  Other terms will of course also be generated, so we cannot say for sure whether ${\cal H}_V$ will have the effects discussed here.
Starting from the $\rSU2$ ASL phase, we find that when the pseudospin breaking is large enough a quantum phase transition to the $\rU1$ ASL results. This phase transition is interesting in its own right as an example of a critical point with no order parameter on either side of the transition,\cite{ran06} and, while we have not pursued this here, it may be fruitful to investigate the critical properties in more detail.  The $\rU1$ ASL can itself undergo an insulator-metal transition (to the semimetal state) upon condensation of a single bosonic field carrying unit charge under the $\rU1$ gauge field.  In the sector with strong pseudospin breaking, we therefore recover the effective theory obtained in the $\rU1$ formulation.\cite{sslee05}

It will now be convenient to introduce some additional notation.
We define the composite field
\begin{equation}
\label{eqn:dbl-cover}
 \big[ {\cal Z}(x) \big]_{i j} = \frac{1}{2} \tr \big( \sigma^i Z^\dagger_x \sigma^j Z_x \big) \text{.}
\end{equation}
It can be shown that ${\cal Z}(x)$ is an ${\rm SO}(3)$ matrix, and that, furthermore, any ${\rm SO}(3)$ matrix can be expressed in this form.  This mapping is two-to-one from $Z$ to ${\cal Z}$ (because $Z$ and $-Z$ map to the same image); in fact, Eq.~(\ref{eqn:dbl-cover})  is nothing but the well-known ``double cover'' map from $\rSU2$ to ${\rm SO}(3)$.  Under an $\rSU2$ gauge transformation sending
$Z_x \to e^{- i \balpha_x \cdot \bsigma / 2} Z_x$,  we have
\begin{equation}
{\cal Z}(x) \to {\cal Z}(x) \big[ R(\balpha_x)\big]^T  \text{,}
\end{equation}
where $R(\balpha_x)$ is the ${\rm SO}(3)$ matrix describing a rotation about the $\balpha_x / | \balpha_x|$ axis by an angle $|\balpha_x|$.  Under a pseudospin rotation $Z_x \to Z_x e^{i \balpha \cdot \bsigma /  2}$, ${\cal Z}$ undergoes a left-${\rm SO}(3)$ rotation:
\begin{equation}
{\cal Z}(x) \to R(\balpha) {\cal Z}(x) \text{.}
\end{equation}

Next, we make use of the double-cover again to define an ${\rm SO}(3)$ gauge field:
\begin{equation}
\big[ {\cal G}^{x x'}\big]_{i j} = \frac{1}{2} \tr \big( U^\dagger_{x x'} \sigma^i U_{x x'} \sigma^j \big) \text{.}
\end{equation}
Under the $\rSU2$ gauge transformation
$U_{x x'} \to e^{-i \balpha_x \cdot \bsigma / 2} U_{x x'} e^{i \balpha_{x'} \cdot \bsigma / 2}$, we have
\begin{equation}
{\cal G}^{x x'} \to R(\balpha_x) {\cal G}^{x x'} \big[R(\balpha_{x'}) \big]^T \text{.}
\end{equation}

The desired pseudospin-breaking term is a gauge-invariant kinetic energy for the composite ${\cal Z}$ boson:
\begin{equation}
\label{eqn:dbl-hopping}
S^{{\rm eff}}_{{\cal Z}} = - t_{c 2} \sum_{\langle x x' \rangle} \tr \Big(R(2 A_{x x'} \bz) {\cal M} \, {\cal Z}(x) {\cal G}^{x x'} \big[{\cal Z}(x')\big]^T  \Big) \text{.}
\end{equation}
Here, ${\cal M}$ is a symmetric $3 \times 3$ matrix, and we have included the coupling to the electromagnetic vector potential $A_{x x'}$ via the rotation matrix $R( 2 A_{x x'} \bz )$.  If we choose ${\cal M} = 1$, then this kinetic term is pseudospin invariant.  Instead, we choose ${\cal M} = \operatorname{diag}(0,0,1)$, which breaks the pseudospin down to ${\rm O}(2)$ as discussed above.  For this choice of ${\cal M}$, the dependence on  $A_{x x'}$ drops out, and we are free to set $A_{x x'} = 0$ in Eq.~(\ref{eqn:dbl-hopping}).  So the only dependence on $A_{x x'}$ is in $S^{{\rm eff}}_Z$ [see Eq.~(\ref{eqn:seff-z-with-emfield})].

Now we consider the effect of making $t_{c 2} \gg 1$, making no assumptions for the moment about the 
other parameters $t_c$ and $K$.
We may proceed by expanding in small fluctuations about the minimum of $S^{{\rm eff}}_{{\cal Z}}$.  We make the following transformation:
\begin{eqnarray}
U_{x x'} &\to& Z_x U_{x x'} Z^\dagger_{x'} \\
F_x &\to& F_x Z^\dagger_x \text{,}
\end{eqnarray}
which also implies
\begin{equation}
{\cal G}^{x x'} \to \big[{\cal Z}(x)\big]^T {\cal G}^{x x'} {\cal Z}(x') \text{.}
\end{equation}
We then have 
\begin{equation}
S^{{\rm eff}}_{\cal Z} = - t_{c2} \sum_{\langle x x' \rangle} \tr \Big( {\cal M} \, {\cal G}^{x x'} \Big) \text{.}
\end{equation}
Choosing ${\cal G}^{x x'}$ (and hence $U_{x x'}$) to minimize $S^{{\rm eff}}_{\cal Z}$, we find that
\begin{equation}
U_{x x'} = \exp \big( i a_{x x'} \sigma^3 \big) \text{.}
\end{equation}

We have thus obtained a Higgs phase in which the $\rSU2$ gauge field has been broken down to $\rU1$.  This corresponds to a condensation of certain components of the composite object ${\cal Z}(x)$.  To better understand the nature of this phase, we work with the remaining terms in the action.  We have
\begin{eqnarray}
S^{{\rm eff}}_Z &=& -2 t_c \sum_{\langle x x' \rangle} \cos \big( A_{x x'} + a_{x x'} \big) \\
S^{{\rm eff}}_F &=& \sum_{\langle x x' \rangle} \tr \Big( F_x e^{i a_{x x'} \sigma^3 } F^\dagger_{x'} \Big) \\
S^{{\rm eff}}_g &=& - 4 K \sum_p \cos \big( [ \nabla \times a]_p \big) \text{.}
\end{eqnarray}
Here, $[ \nabla \times a]_p$ is the lattice curl of $a_{x x'}$, taken around the perimeter of the plaquette $p$.  Next, we make the transformation $a_{x x'} \to a_{x x'} + \theta_x - \theta_{x'}$ and also $F_x \to F_x e^{- i \theta_x \sigma^3}$; this introduces the bosonic matter field $\theta_x$ and restores a $\rU1$ gauge invariance.  The form of $S^{{\rm eff}}_F$ and $S^{{\rm eff}}_g$ is unchanged, and we have
\begin{equation}
S^{{\rm eff}}_Z = -2 t_c \sum_{\langle x x' \rangle} \cos \big( \theta_x - \theta_{x'} + A_{x x'} + a_{x x'} \big)
\text{.}
\end{equation}
We have thus arrived at the effective action obtained in the $\rU1$ slave-rotor formulation.  The compact $\rU1$ gauge field $a_{x x'}$ is coupled to the charge-neutral, fermionic spinons and also the rotor boson field with phase variable $\theta_x$.  The $\theta$-boson carries unit electromagnetic charge, as is apparent from the presence of the external electromagnetic field in $S^{{\rm eff}}_Z$.

We now consider the consequences of the above result for the phase diagram, focusing on the regime $K \gg 1$.  It is useful to imagine starting in the $\rSU2$ ASL phase, where $t_c, t_{c2} \ll 1$. As $t_{c 2}$ is increased, eventually ${\cal Z}$ will condense and the $\rU1$ gauge theory action given above is appropriate.  Because $t_c \ll 1$, the $\theta$-boson is gapped, so the resulting phase is the $\rU1$ ASL.  The quantum critical point where ${\cal Z}$ condenses is thus a phase transition between the $\rSU2$ and $\rU1$ algebraic spin liquids.  Next, we keep $t_{c 2}$ large and increase $t_c$.  Eventually the $\theta$-boson will condense, leading to the semimetal.\cite{sslee05}  So, in the regime $t_{c2} \gg 1$, we recover the scenario of Ref.~\onlinecite{sslee05}.  Also, it should be noted that the semimetal phase persists into the regime $t_c \gg 1$ and $t_{c2} \ll 1$.

\begin{figure}
\includegraphics[width=3in]{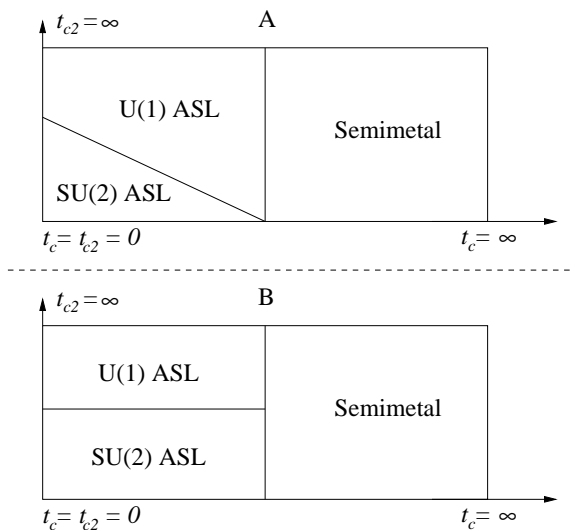}
\caption{Schematic phase diagrams of the effective action in the $t_c$-$t_{c2}$ plane; $K$ is taken to be large.  As discussed in the text, two scenarios are shown, with differing assumptions about the critical behavior at the $\rSU2$ ASL to semimetal transition, and hence different topologies of the phase diagram.}
\label{fig:psbreaking}
\end{figure}

Schematic phase diagrams reflecting these considerations are shown in Fig.~\ref{fig:psbreaking}.  Depending on the (unknown) critical properties of the $\rSU2$ ASL-semimetal transition, different topologies of the phase diagram are possible; we discuss the two simplest scenarios.
 In Fig.~\ref{fig:psbreaking}A, pseudospin breaking is assumed to be relevant; in this scenario,  the Mott transition for $t_{c2} \neq 0$ is always the $\rU1$ ASL-semimetal transition.  On the other hand, in Fig.~\ref{fig:psbreaking}B, pseudospin breaking is taken to be irrelevant, and $t_{c 2}$ must be increased above a critical value to change the nature of the Mott transition.  Other, more complicated scenarios  are of course also possible.  It should be noted that, starting from within the $\rSU2$ ASL phase, $t_{c2}$ must always be increased above a critical value to reach the phase transition to the $\rU1$ ASL; this follows from the $Z$-boson gap in the $\rSU2$ ASL.

\section{Discussion}
\label{sec:discussion}

The $\rSU2$ slave-rotor formulation developed in this paper is a general formalism to describe spin liquids and related physics in Hubbard models; it is especially useful at half-filling to describe spin liquid ground states near the Mott transition.  We proposed the honeycomb Hubbard model as a candidate for this physics, and showed that the simplest possibility is a direct, continuous quantum phase transition between the conducting semimetal and an $\rSU2$ algebraic spin liquid.  

We hope that this physics will be looked for in future quantum Monte Carlo simulations of the honeycomb model; to that end we 
summarize our main predictions and suggestions for such work.  First, and most simply, the mean-field analysis suggests that the spin liquid phase is more likely to be found if a third-neighbor electron hopping $t''$ is present, when $t'' / t < 0$.  Second, a robust signature of the $\rSU2$ ASL is crucial for its detection in numerical simulations, especially as one may be limited to rather small systems, and only a small region of spin liquid phase may be present.   The simultaneous presence of slowly decaying power-law N\'{e}el and valence-bond solid correlations, with the same critical exponent, should provide such a signature.  The particular pattern of VBS order with slowly decaying correlations is shown in Fig.~\ref{fig:hcvbs}, and some specific examples of correlation functions exhibiting the slow power-law decay are given in Sec.~\ref{sec:spin-liquid-ft}.  The VBS correlations can be measured either using a dimer-dimer correlator, or by the correlation function of the bond kinetic energy.  While limitation to small system sizes may prevent a detailed analysis of the critical behavior, it should be possible to look for the simultaneous presence of slowly-decaying N\'{e}el and VBS correlations, which would already be some evidence for the $\rSU2$ ASL.

Finally, it may be useful to consider an extended Hubbard model with the nearest-neighbor density interaction ${\cal H}_V$ of Eq.~(\ref{eqn:hv-term}).  This term breaks the pseudospin symmetry down to ${\rm O}(2)$, and, as discussed in Sec.~\ref{sec:u1sl}, it may favor a $\rU1$ ASL over the $\rSU2$ ASL.  This is potentially advantageous because the $\rU1$ ASL is likely more stable against small perturbations than its $\rSU2$ counterpart, because gauge theories with more gauge bosons (\emph{i.e.} more colors) have a stronger tendency toward confinement.  

Throughout this paper we have been assuming that the $\rSU2$ ASL is a stable phase; while this does hold in the large-$N_f$ limit, it may not be true in the physical case of the Hubbard model.  One possibility is that the $\rSU2$ ASL is not a stable phase, but its $\rU1$ counterpart is, and in that case a pseudospin breaking term such as ${\cal H}_V$ may be required to obtain a stable spin liquid adjacent to the Mott transition in the honeycomb Hubbard model.  (We caution that the $\rU1$ ASL itself is also not guaranteed to be a stable phase.)

As in the $\rSU2$ case, the $\rU1$ ASL can be tested for by looking for competing orders with slowly decaying power-law correlations.  These competing orders are unified together by an ${\rm SU}(4)$ symmetry, and their correlations fall off with the same critical exponent -- the value of this exponent will be different from its counterpart in the $\rSU2$ ASL.  The pattern of competing orders could be determined in a straightforward fashion following Ref.~\onlinecite{hermele05}, and is a superset of those found in the $\rSU2$ ASL, including the N\'{e}el and VBS correlations discussed here, as well as correlations in other observables.

For the case of nearest-neighbor hopping only, the honeycomb Hubbard model has actually already been studied via quantum Monte Carlo and series expansion techniques.\cite{paiva05}  Those results suggest, but do not conclusively establish, a direct, continuous quantum phase transition between the SM and AF states for $(U/t)_c \approx 4 - 5$, which could be the conventional SM-AF transition\cite{herbut06} shown in Fig.~\ref{fig:scenarios}a.  Of course, a small region of spin liquid intervening between SM and AF is not excluded.  Another possibility, if the $\rSU2$ ASL is not a stable phase, is a deconfined critical point\cite{senthil04a, senthil04b} from SM to AF.  Briefly, this can be understood as a transition from SM to the $\rSU2$ ASL, which is in turn unstable (via a dangerously irrelevant perturbation) to N\'{e}el order.  
In order to obtain some information about which scenario is in fact occurring, the dimer-dimer or bond kinetic energy correlation functions could be measured to look for an enhanced tendency toward VBS ordering near the transition.  Also, as discussed above, adding a $t''$ hopping or the nearest-neighbor repulsion ${\cal H}_V$ may help to stabilize a spin liquid phase or deconfined critical point.

While our main focus in this paper has been the honeycomb lattice, the $\rSU2$ slave-rotor formulation should be useful much more generally, as an extension of the $\rSU2$ gauge theory of the Heisenberg model to the case of the Hubbard model.  It would clearly be interesting to investigate the triangular lattice using this approach, motivated by experiments on $\kappa$-(BEDT-TTF)$_2$Cu$_2$(CN)$_3$.  This would allow the competition among a wider range of spin liquids than that considered in Ref.~\onlinecite{sslee05} to be addressed.  The simplest way to carry out such an analysis is via mean-field theory, but a study of variational wavefunctions is probably more reliable.  It may be possible to use the $\rSU2$ slave-rotor formulation to develop new variational wavefunctions for Hubbard models, to study physics near the Mott transition.

On an even more general note, the $\rSU2$ slave-rotor formulation may help to shed more light on the physical nature of fermionic spinons.  For most spin liquids with fermionic spinons, we lack a simple physical picture of the spinons in terms of microscopic degrees of freedom.  One can draw the usual cartoon pictures of resonating valence-bond states, and show that breaking a valence bond corresponds to creating a pair of spinons; however, such pictures do nothing to distinguish among the great variety of different spin liquid states, and do not even tell us whether the spinons should be thought of as fermions or bosons.  

It turns out that more progress has been made in coming to a physical understanding of bosonic spinons.  The reason for this is that spin liquids with bosonic spinons can be thought of as adjacent to a \emph{magnetically ordered} state, obtained upon condensation of the spinons.  In some cases, the spin liquid phase, and hence the spinons, can then be understood in terms of the degrees of freedom of the ordered state.  For example, in two-dimensional $S = 1/2$ systems with XY symmetry, a $Z_2$ spin liquid be obtained upon condensation of pairs of vortices in the XY-ordered state.\cite{senthil00}  The deconfined critical point between the VBS and the XY-ordered state, and its bosonic spinons, can also be understood in terms of the vortices of the XY-ordered state.\cite{senthil04a, senthil04b}

The $\rSU2$ slave-rotor formulation suggests a similar viewpoint for spin liquids with fermionic spinons, as it should allow for a description of the Mott transition out of \emph{any} such spin liquid, into a conventional conducting state (either a metal or superconductor).  At the transition, the fermionic spinons transform into electron-like quasiparticles.  This suggests that a physical picture of the spinons might be obtained by understanding the spin liquid in terms of the degrees of freedom of the nearby conducting state.  This may be a challenging task, but it has been carried out in one case:  starting from a $d$-wave superconductor with gapless nodal quasiparticles, condensation of pairs of superconducting vortices leads to a $Z_2$ spin liquid with nodal fermionic spinons.\cite{senthil00, balents98,balents99,balents00}  We are optimistic that further progress along similar lines is possible in other cases.

\begin{acknowledgments}
We have benefited greatly from discussions with Sung-Sik Lee, Daniel Podolsky and Ying Ran.  We are especially grateful to Patrick Lee for suggesting the problem of the honeycomb lattice model, and also for numerous discussions and a careful reading of the manuscript.  This work is supported by funding from the NEC Corporation, NSF Grant No. DMR-0433632, and the Department of Energy under grant DE-FG02-03ER46076.
\end{acknowledgments}

\appendix
\section{Quantum mechanics of the $\rSU2$ matrix rotor}
\label{app:su2rotor}

In this appendix we shall derive some details of the quantum mechanics of the $\rSU2$ matrix rotor.  In particular, we develop its functional integral representation, both with and without an $\rSU2$ gauge constraint.  As a warmup we first review the ${\rm O}(3)$ quantum rotor, then, in the following section we return to the $\rSU2$ matrix rotor.

\subsection{${\rm O}(3)$ quantum rotor}

Consider a particle constrained to move on the surface of the 2-sphere $S^2$.  Its position is given by the unit-length 3-vector $\bx$  (\emph{i.e.} $\bx^2 = 1$).  The quantum Hilbert space has a basis of position eigenstates $\{ | \bx \rangle \}$, with the usual inner product $\langle \bx | \bx' \rangle = \delta(\bx - \bx')$.  Note that this is a ``two-dimensional'' delta function, as appropriate for a variable constrained to a 2-dimensional manifold, so, for example, $\int_{S^2} d\bx \, \delta(\bx) = 1$.

Angular momentum operators $\bL$ are defined as the generators of infinitesimal rotations.  That is, for a finite rotation, we have
\begin{equation}
e^{i \balpha \cdot \bL } | \bx \rangle = | R(\balpha) \bx \rangle \text{,}
\end{equation}
where $R(\balpha)$ denotes a rotation by $|\balpha|$ about the $\balpha / |\balpha|$ axis.  This implies the familiar commutation relations $[ L^i, L^j] = i \epsilon^{i j k} L^k$ and $[ L^i, x^j] = i \epsilon^{i j k} x^k$.

Next, we construct a basis of angular momentum eigenstates.  First of all, there is a unique singlet state given by
\begin{equation}
|0\rangle = \frac{1}{\sqrt{4\pi}} \int_{S^2} d\bx\, | \bx \rangle \text{.}
\end{equation}
The higher angular momentum eigenstates are spherical harmonics, which can be obtained by acting on $| 0 \rangle$ with irreducible polynomials of $x_i$; that is, polynomials that are symmetric in ${\rm O}(3)$ vector indices and traceless in any pair of two such indices.  Such polynomials transform under irreducible representations of the rotation group, hence the name.  For example, we have, not worrying about normalization,
\begin{eqnarray}
| i \rangle &=& x_i | 0 \rangle \\
| i j \rangle &=& (x_i x_j - \frac{1}{3} \delta_{i j} ) | 0 \rangle \\
| i j k \rangle &=& \big[ x_i x_j x_k - \frac{1}{5} ( \delta_{i j} x_k + \delta_{i k} x_j + \delta_{j k} x_i ) \big]
| 0 \rangle
\text{.}
\end{eqnarray}
These states transform with total angular momentum $\ell = 1,2$ and $3$, respectively.  Including arbitrary high values of $\ell$, the angular momentum eigenstates form a basis for the Hilbert space.

Consider the Hamiltonian
\begin{equation}
H = \frac{I}{2} \bL^2 \text{.}
\end{equation}
We would like to go to a functional integral description of the partition function.  In order to do this, we will need to evaluate objects like
\begin{equation}
\langle \bx(\tau + \epsilon) | e^{-\epsilon H} | \bx(\tau) \rangle \text{.}
\end{equation}
As usual we assume only smooth configurations of $\bx(\tau)$ contribute to the path integral, so that
$\bx(\tau + \epsilon) = \bx(\tau) + {\cal O}(\epsilon)$.  It should be enough to consider only the first-order term in $\epsilon$, which will dominate the ${\cal O}(\epsilon^2)$ and higher terms.  Then, to first order in $\epsilon$, we note that $\bx(\tau + \epsilon)$ lies in the plane tangent to the sphere at $\bx(\tau)$.  So, for the purposes of evaluating this matrix element, we can \emph{change the Hilbert space} to that of a particle moving in this tangent plane.  So we replace $\bx \to \bx_T$, where $\bx_T$ lies in the plane tangent to the sphere at $\bx(\tau)$, and $\bx_T(\tau) = \bx(\tau)$.
We then have
\begin{equation}
\langle \bx(\tau + \epsilon) | e^{-\epsilon H} | \bx(\tau) \rangle
\rightarrow
\langle \bx_T(\tau + \epsilon) | e^{-\epsilon H_T} | \bx_T(\tau) \rangle \text{.}
\end{equation}

In order to complete this we have to understand how to define $H_T$, the Hamiltonian in the tangent-plane Hilbert space.  This is seen most easily by considering a concrete example, where $\bx(\tau) = \boldsymbol{z}$.  Then we see that
\begin{eqnarray}
e^{i \theta L^z} | \boldsymbol{z} \rangle &=& | \boldsymbol{z} \rangle \\
e^{i \theta L^x} | \boldsymbol{z} \rangle &=& | \boldsymbol{z} - \theta \boldsymbol{y} + {\cal O}(\theta^2) \rangle \\
e^{i \theta L^y} | \boldsymbol{z} \rangle &=& | \boldsymbol{z} + \theta \boldsymbol{x} + {\cal O}(\theta^2) \rangle \text{.}
\end{eqnarray} 
For infinitesimal rotations, the angular momentum operators generate \emph{translations} in the tangent plane, and so should be replaced with the linear momentum operators in the tangent plane Hilbert space.  Specifically we have
\begin{eqnarray}
L_x &\to& - p_y \\
L_y &\to& p_x \\
L_z &\to& 0 \text{.}
\end{eqnarray}
More generally, if we wish to consider the tangent plane at an arbitrary point $\bx_0$, we have
\begin{equation}
\bL \to \bx_0 \times \boldsymbol{p}_\perp \text{,}
\end{equation}
where $\boldsymbol{p}_\perp$ is the linear momentum in the tangent plane, and $\boldsymbol{p}_\perp \cdot \bx_0 = 0$.  Under this replacement, note that we have simply $\bL^2 \to \boldsymbol{p}_\perp^2$,
so we have $H_T = \frac{I}{2} \boldsymbol{p}_\perp^2$.

Thus we can write
\begin{eqnarray}
&& \langle \bx(\tau + \epsilon) | e^{-\epsilon H} | \bx(\tau) \rangle \approx
\langle \bx_T(\tau + \epsilon) | e^{-\epsilon H_T} | \bx_T(\tau) \rangle \nonumber \\
&=&
\int d^2 \boldsymbol{p}_\perp \langle \bx_T(\tau + \epsilon) | \boldsymbol{p}_\perp \rangle
e^{- \frac{\epsilon I}{2} \boldsymbol{p}_\perp^2 } \langle \boldsymbol{p}_\perp | \bx_T(\tau) \rangle \text{.}
\end{eqnarray}
It is now clear that we will arrive at the functional integral:
\begin{equation}
{\cal Z} = \operatorname{Tr} e^{-\beta H} 
= \int {\cal D} \bx(\tau) \int {\cal D} \boldsymbol{p}_\perp(\tau, \bx(\tau)) e^{- S[ \bx, \boldsymbol{p}_{\perp} ]} \text{,}
\end{equation}
where $\bx$ is constrained to lie on the unit sphere, it should be noted that the range of integration of $\boldsymbol{p}_\perp$ depends on $\bx(\tau)$, and the action has the form
\begin{equation}
S[\bx, \bpperp] = \int_0^\beta d\tau \Big( - i \bpperp \cdot \dot{\bx} + \frac{I}{2} \bpperp^2 \Big) \text{.}
\end{equation}
Upon integrating out the momentum variable we obtain the familiar form
\begin{equation}
S[\bx] = \frac{1}{2 I}  \int_0^\beta d\tau (\partial_\tau \bx)^2 \text{.}
\end{equation}

\subsection{$\rSU2$ matrix rotor}

Now that we have dispensed with our warmup, let us consider the situation we need for the analysis in the paper, the quantum mechanics of the $\rSU2$ matrix rotor.  As mentioned in Sec.~\ref{sec:su2-formulation}, this is identical to a particle constrained to the 3-sphere with ${\rm O}(4)$ symmetry, and here we shall find it convenient to work with the ${\rm O}(4)$ rotor description, especially to derive the functional integral.  The position of the particle is specified by the $\rSU2$ matrix $Z$, or equivalently by the unit-length 4-vector $x_{\mu}$, with $\mu = 0,\dots,3$.  The equivalence of these two representations if manifest upon writing
\begin{equation}
Z = x_0 - i \bsigma \cdot \bx \text{,}
\end{equation}
where $\bx = (x_1, x_2, x_3)$.  The Hilbert space is specified by the basis of position eigenstates $\{ | Z \rangle \}$ or, equivalently, $\{ | x \rangle \}$.

${\rm SO}(4)$ rotations of the position vector $x_\mu$ are generated by the operators $L_{\mu \nu} = - L_{\nu \mu}$, which satisfy the commutation relations
\begin{equation}
{[} L_{\mu \nu}, L_{\lambda \sigma} {]}
= i ( \delta_{\mu \lambda} L_{\nu \sigma}
+ \delta_{\nu\sigma} L_{\mu \lambda} 
- \delta_{\mu \sigma} L_{\nu \lambda}
- \delta_{\nu \lambda} L_{\mu \sigma} ) \text{,}
\end{equation}
and
\begin{equation}
{[} L_{\mu \nu}, x_\lambda {]} = i ( \delta_{\mu \lambda} x_{\nu} - \delta_{\nu \lambda} x_{\mu} ) \text{.}
\end{equation}
Now, ${\rm SO}(4) = \rSU2 \times \rSU2 / {\rm Z}_2$, and we can express the generators of right and left $\rSU2$ rotations of $Z$, $\bJ_R$ and $\bJ_L$, in terms of the $L_{\mu \nu}$.  To see this, we follow the usual procedure and define
\begin{eqnarray}
J_i &=& \frac{1}{2} \epsilon_{i j k} L_{j k} \\
K_i &=& L_{0 i} \text{,}
\end{eqnarray}
where $i = 1,\dots,3$.  In terms of these operators
\begin{eqnarray}
\bJ_R &=& \frac{1}{2} ( \bJ - \bK ) \\
\bJ_L &=& \frac{1}{2} ( \bJ + \bK ) \text{,}
\end{eqnarray}
which satisfy the commutation relations and other identities given in Sec.~\ref{sec:su2-formulation}.

As for the ${\rm O}(3)$ rotor, we need to define a basis of angular momentum eigenstates.  There is again a unique rotationally-invariant state
\begin{equation}
|0\rangle_{{\rm rot}} = \frac{1}{\sqrt{2\pi^2}} \int_{S^3} d^3 x \, | x \rangle \text{.}
\end{equation}
By analogy with the ${\rm O}(3)$ rotor, we should be able to access all higher angular momentum states by acting on $|0 \rangle$ with irreducible polynomials in $x_\mu$.  Again, such polynomials are completely symmetric in ${\rm O}(4)$ vector indices, traceless over any pair of indices, and therefore transform as irreducible representations of ${\rm O}(4)$.  For example, $x_\mu$ and $[x_\mu x_\nu - (1/4) \delta_{\mu \nu}]$ satisfy these conditions.  For the irreducible polynomial of degree $k$, there are $\operatorname{dim}_k$ independent components.  A simple counting argument shows
\begin{equation}
\label{eqn:counting}
\operatorname{dim}_k = (k+1)^2 \text{.}
\end{equation}

Because they transform as irreducible representations of the ${\rm SO}(4)$ and hence $\rSU2 \times \rSU2$ Lie algebra, the states just constructed can be labeled by the quantum numbers $(\ell_R, \ell_L)$, the total angular momentum quantum numbers under right and left rotations, respectively.  Only some pairs  $(\ell_R, \ell_L)$ actually appear in the Hilbert space -- we will show that the multiplets with $\ell_R = \ell_L$ form a complete basis.  Now, we observe that we can construct the representation $(\ell, \ell)$ in our Hilbert space, for $\ell = 0, 1/2, 1, 3/2, \dots$.  This is done most easily in terms of $Z$, which we can write as a two-index tensor $Z_{a \alpha}$, where I use roman letters for the first index and Greek letters for the second one because they transform under independent $\rSU2$'s.  The state corresponding to $(0, 0)$ is just $|0\rangle_{{\rm rot}}$, and clearly the $(1/2, 1/2)$ multiplet is made up of the states $Z_{a \alpha} | 0 \rangle_{{\rm rot}}$.  In general, forgetting about normalization, we have
\begin{equation}
\label{eqn:zstates}
(n/2, n/2) \leftrightarrow \sum_P Z_{a_{P(1)} \alpha_1} \cdots Z_{a_{P(n)} \alpha_n}  |0 \rangle_{{\rm rot}}
\text{,}
\end{equation}
where the sum is over all permutations of $\{1,\dots, n \}$.
These states are completely symmetric in both the $a_i$ and $\alpha_i$ indices, which implies that they transform as a $(n/2, n/2)$ multiplet.  Now, the $(n/2, n/2)$ multiplet contains $(n+1)^2$ states, so we are forced to conclude that the states of Eq.~(\ref{eqn:zstates}) correspond to those obtained from the irreducible polynomial of degree $n$.  This exhausts all possible states, so the $(\ell, \ell)$ multiplets form a basis.

Next we will construct a functional integral, first for the Hamiltonian
\begin{equation}
H = \frac{I}{2} \sum_{\mu < \nu} L^2_{\mu \nu} = I ( \bJ_R^2 + \bJ_L^2 ) \text{.}
\end{equation}
Again, to evaluate each matrix element in the Trotter expansion, we need to replace the Hilbert space of the 3-sphere by the Hilbert space of the tangent hyperspace.  So, for a particle with position $x$, we make the replacement $L_{\mu \nu} \to x_\mu p_\nu - x_\nu p_\mu$, where $p_\mu x_\mu = 0$.  Under this replacement we have $\sum_{\mu < \nu} L^2_{\mu \nu} \to p_\mu p_\mu$.  It is then clear that the functional integral takes the same form as before, but in one higher dimension:
\begin{eqnarray}
{\cal Z} &=& \int {\cal D}x(\tau) \int {\cal D}p(\tau, x(\tau)) e^{-S[x, p]} \\
S[x,p] &=& \int_0^\beta d\tau \Big( - i p_\mu \dot{x}_\mu + \frac{I}{2} p_\mu p_\mu \Big ) \text{.}
\end{eqnarray}
Again, upon integrating out the momentum variable, we have
\begin{equation}
S[x] = \frac{1}{2 I} \int_0^\beta d\tau (\partial_\tau x_\mu )^2 \text{.}
\end{equation}

Let us rewrite this in $\rSU2$ matrix notation.  We have
\begin{eqnarray}
&& \tr (\partial_\tau Z^\dagger \partial_\tau Z ) \nonumber \\
 &=& \tr \Big[ (\partial_\tau x_0 + i \bsigma \cdot \partial_\tau \bx ) (\partial_\tau x_0 - i \bsigma \cdot \partial_\tau \bx ) \Big] \nonumber \\
&=& 2 (\partial_\tau x_\mu )^2 \text{,}
\end{eqnarray}
and therefore
\begin{eqnarray}
\label{eqn:zpf}
{\cal Z} &=& \int {\cal D}Z(\tau) e^{- S[Z] } \\
S[Z] &=& \frac{1}{4 I} \int_0^\beta d\tau \tr (\partial_\tau Z^\dagger \partial_\tau Z) \text{.}
\end{eqnarray}
Here, $Z(\tau)$ is constrained to be an $\rSU2$ matrix.

Now let us consider a different Hamiltonian problem, which is exactly what we will need to write down the functional integral form of the Hubbard model in the $\rSU2$ slave rotor formulation.  Suppose that
$H = I \bJ_R^2$, and there is a constraint involving $\bJ_L$:
\begin{equation}
\bJ_L + \boldsymbol{v} = 0 \text{.}
\end{equation}
Here, $\boldsymbol{v}$ is a vector that may depend on other dynamical fields (\emph{e.g.} the fermionic spinons) but commutes with all operators acting in the rotor Hilbert space.  So for the present considerations we can take it to be a $c$-number.  We will define a functional integral by tracing over the unconstrained Hilbert space, but we define a projection operator ${\cal P}$ that projects onto the constrained Hilbert space.  Then we have
\begin{equation}
{\cal Z} = \tr ({\cal P} e^{-\beta H} ) = \tr ({\cal P} e^{-\epsilon H} {\cal P} e^{-\epsilon H} \dots 
{\cal P} e^{-\epsilon H} ) \text{,}
\end{equation}
which holds because $[{\cal P}, H] = 0$.  Now we have
\begin{equation}
{\cal P} \propto \int d\boldsymbol{\lambda} \, e^{- i \epsilon \boldsymbol{\lambda} \cdot (\bJ_L +\boldsymbol{v} )} \text{,}
\end{equation}
where the unimportant constant of proportionality will be absorbed into the functional integration measure.  We now see that we need to evaluate objects like
\begin{equation}
\langle x(\tau + \epsilon) | e^{- i \epsilon \boldsymbol{\lambda} \cdot (\bJ_L +\boldsymbol{v} )} e^{-\epsilon H} | x(\tau) \rangle \text{,}
\end{equation}
which can be evaluated precisely as above by going to the tangent hyperplane Hilbert space.  It can be shown that the correct replacement for $\bJ_R^2$ is
\begin{equation}
\bJ_R^2 \to  \frac{1}{4} p_\mu p_\mu \text{.}
\end{equation}
We then obtain
\begin{equation}
{\cal Z} = \int {\cal D} x(\tau) {\cal D} \boldsymbol{\lambda}(\tau) \int {\cal D} p(\tau, x(\tau)) e^{-S[x,p,\blambda]} \text{,}
\end{equation}
and
\begin{eqnarray}
S[x,p,\blambda] &=& \int_0^\beta d\tau \Big(
- i p_\mu \dot{x}_\mu + \frac{I}{4} p_\mu p_\mu \\ &+& \frac{i}{2} \blambda \cdot ( \bx \times \bp + x_0 \bp - \bx p_0) +
i \blambda \cdot \boldsymbol{v} \Big) \text{.} \nonumber
\end{eqnarray}
Upon integrating out the momentum, we obtain
\begin{eqnarray}
S[x, \blambda] &=& \int_0^\beta d\tau \Big[ \frac{1}{I} (\partial_\tau x_0 + \frac{1}{2} \blambda \cdot \bx)^2
\nonumber \\ &+& \frac{1}{I} (\partial_\tau \bx - \frac{1}{2} \lambda \times \bx - \frac{1}{2} x_0 \blambda)^2 + i \blambda \cdot \boldsymbol{v} \Big] \\
&=& \frac{1}{2 I} \int_0^\beta d\tau \Big\{ \tr \Big[ Z^\dagger (\overleftarrow{\partial}_\tau - \frac{i \blambda \cdot \bsigma}{2} ) ( \partial_\tau + \frac{i \blambda \cdot \bsigma}{2} ) Z \Big]  \nonumber \\ &+& i \blambda \cdot \boldsymbol{v} \Big\} \text{,}
\end{eqnarray}
where the second equality can be verified by simple algebra after expressing $Z = x_0 - i \bsigma \cdot \bx$.

\section{$\rSU2$ ASL Field Theory}
\label{app:field-theory}

Here, we shall develop some of the details of the field theory of the $\rSU2$ algebraic spin liquid in the honeycomb Hubbard model.  We first work through the continuum limit, and write the Lagrangian in a form that manifestly exposes the ${\rm Sp}(4)$ symmetry discovered by Ran and Wen.\cite{ran06}  Next, we enumerate the action of the microscopic symmetries on the fields of the low-energy theory.  Finally, we discuss the 5-component vector of ``competing orders,'' in which the N\'{e}el vector and two-component VBS order parameter are unified together by the ${\rm Sp}(4)$ symmetry.

\subsection{Continuum limit and ${\rm Sp}(4)$ symmetry}

Consider the mean-field Hamiltonian
\begin{equation}
{\cal H}_F = t_f \sum_{\langle \br \br' \rangle} \big( \psi^\dagger_{\br} \psi^{\vphantom\dagger}_{\br'} + \text{H.c.} \big) \text{,}
\end{equation}
where the sum is over near-neighbor honeycomb bonds, and $\psi_{\br}$, which carries the $\rSU2$ gauge charge, is defined in Eq.~(\ref{eqn:psi-defn}).
We will take the continuum limit and catalog the action of the microscopic symmetries on the continuum fields.

As in Sec.~\ref{sec:intro} and Fig.~\ref{fig:honeycomb},  we use the 2-site unit cell labeled by $(\bR,i)$, with $i = 1,2$, and define the Fourier transform
\begin{equation}
\psi_{\bR i} = \frac{1}{\sqrt{N_c}} \sum_{\bk} e^{i \bk \cdot \bR} \psi_{\bk i} \text{,}
\end{equation}
where $N_c$ is the number of unit cells in the lattice.  The Hamiltonian then becomes
\begin{equation}
{\cal H}_F = t_f \sum_{\bk} 
\left( \begin{array}{cc}
\psi^\dagger_{\bk 1} & \psi^\dagger_{\bk 2}
\end{array} \right)
H(\bk) 
\left( \begin{array}{c}
\psi_{\bk 1} \\
\psi_{\bk 2}
\end{array} \right) \text{,}
\end{equation}
where 
\begin{equation}
\left[H(\bk)\right]_{1 1} = \left[H(\bk)\right]_{2 2} = 0 \text{,}
\end{equation}
and
\begin{equation}
\left[H(\bk)\right]_{1 2} = \left[H(\bk)\right]^*_{2 1} = 1 + e^{-i \bk \cdot \ba_2} + e^{i \bk \cdot (\ba_1 - \ba_2)} \text{.}
\end{equation}

The Dirac nodes are at $\bk = \pm \bQ$, where $\bQ = (4\pi/3)\bx$.  Putting $\bk = \pm \bQ + \bq$ and expanding to first order in small $\bq$ we find
\begin{equation}
H(\bQ + \bq) = - \frac{\sqrt{3}}{2} (q_x \tau^1 + q_y \tau^2) \text{,}
\end{equation}
and
\begin{equation}
H(-\bQ + \bq) = \frac{\sqrt{3}}{2} (q_x \tau^1 - q_y \tau^2 ) \text{.}
\end{equation}
We define the continuum spinors
\begin{equation}
\tilde{\varphi}_1(\bq) \sim \left( \begin{array}{c}
\psi_{\bQ + \bq, 1} \\
\psi_{\bQ + \bq, 2}
\end{array} \right) \text{,}
\end{equation}
and
\begin{equation}
\tilde{\varphi}_2(\bq) \sim \left( \begin{array}{c}
\psi_{-\bQ + \bq, 1} \\
\psi_{-\bQ + \bq, 2}
\end{array} \right) \text{,}
\end{equation}
Note that $\tilde{\varphi}_a$ (with $a = 1,2$) is a 4-component object.  As discussed in Sec.~\ref{sec:spin-liquid-ft} it is convenient to act on this space with tensor products of two different kinds of Pauli matrices.  The $\tau^i$ Pauli matrices act on the Lorentz space, and the $\mu^i$ Pauli matrices act in the $\rSU2$ gauge space.

The continuum Hamiltonian takes the form
\begin{equation}
{\cal H}_{c} = v \int \frac{d^2\bq}{(2\pi)^2}
\Big[ \tilde{\varphi}^\dagger_1 (-q_x \tau^1 - q_y \tau^2) \tilde{\varphi}_1
+ \tilde{\varphi}^\dagger_2 (q_x \tau^1 - q_y \tau^2) \tilde{\varphi}_2 \Big] \text{,}
\end{equation}
where $v$ is the velocity.  To make this look like the conventional Dirac Hamiltonian we define
\begin{eqnarray}
\varphi_1 &\equiv& \tau^3 \tilde{\varphi}_1 \\
\varphi_2 &\equiv& \tau^1 \tilde{\varphi}_2 \text{,}
\end{eqnarray}
and we have
\begin{equation}
{\cal H}_{c} = v \int \frac{d^2\bq}{(2\pi)^2} \varphi^\dagger_a (q_x \tau^1 + q_y \tau^2) \varphi^{\vphantom\dagger}_a \text{.}
\end{equation}

It is useful to write down again the relation between the continuum and lattice fields:
\begin{equation}
\label{eqn:varphi1}
\varphi_1(\bq) \sim \left( \begin{array}{c}
\psi_{\bQ + \bq, 1} \\
- \psi_{\bQ + \bq, 2}
\end{array} \right) \text{,}
\end{equation}
and
\begin{equation}
\label{eqn:varphi2}
\varphi_2(\bq) \sim \left( \begin{array}{c}
\psi_{-\bQ + \bq, 2} \\
\psi_{-\bQ + \bq, 1} 
\end{array} \right) \text{.}
\end{equation}
We also define the eight-component object
\begin{equation}
\Phi = \left( \begin{array}{c}
\varphi_1 \\
\varphi_2 
\end{array} \right) \text{.}
\end{equation}
This introduces an additional $\rSU2$ \emph{flavor} space -- we denote the Pauli matrices acting there by $\nu^i$.  So, for example,
\begin{equation}
\nu^1 \Phi = \left( \begin{array}{c} \varphi_2 \\ \varphi_1 \end{array} \right) \text{.}
\end{equation}

We now write down the imaginary-time Lagrangian density
\begin{eqnarray}
{\cal L}_0 &=& \varphi^\dagger_a \Big[ \partial_0 - i  (\tau^1 \partial_1 + \tau^2 \partial_2) \Big] \varphi_a \\
&=& \bar{\varphi}_a \big[ -i \gamma_\mu \partial_\mu \big] \varphi_a \\
&=& \bar{\Phi} \big[ -i \gamma_\mu \partial_\mu \big] \Phi \text{,}
\end{eqnarray}
where we have set $v=1$ for convenience, and defined
\begin{equation}
\gamma_\mu = (\tau^3, \tau^2, -\tau^1) \text{,}
\end{equation}
and
\begin{eqnarray}
\bar{\varphi}_a &=& i \varphi^\dagger_a \tau^3 \\
\bar{\Phi} &=& i \Phi^\dagger \tau^3 \text{.}
\end{eqnarray}

The coupling to the $\rSU2$ gauge field $a^i_{\mu}$ is introduced by replacing $\partial_\mu$ with the covariant derivative
\begin{equation}
D_\mu \equiv \partial_\mu + \frac{i a^i_\mu \mu^i}{2} \text{,}
\end{equation}
resulting in the fermion kinetic energy part of the effective Lagrangian density
\begin{equation}
{\cal L}_{{\rm eff}} = \bar{\Phi} \big[ -i \gamma_\mu D_\mu \big] \Phi
= \bar{\Phi} \big[ -i \Dslash \big] \Phi \text{,}
\end{equation}
where we have introduced the Feynman ``slash'' notation $\Dslash \equiv \gamma_\mu D_\mu$.

We now proceed to put ${\cal L}_{{\rm eff}}$ into a form that manifestly exposes the ${\rm Sp}(4)$ symmetry of Ran and Wen.\cite{ran06}  First, we define another kind of complex conjugate spinor
\begin{equation}
\hat{\Phi} = (i\mu^2)(i\tau^2) \bar{\Phi}^T \text{.}
\end{equation}
The form of $\hat{\Phi}$ takes advantage of the fact that the spin-$1/2$ representation of $\rSU2$ is self-conjugate, so that $\Phi$ and $\hat{\Phi}$ transform identically under both Lorentz rotations and gauge transformations.  For example, under a Lorentz rotation, $\Phi \to e^{i \alpha_\mu \gamma_\mu} \Phi$, and also $\hat{\Phi} \to e^{i \alpha_\mu \gamma_\mu} \hat{\Phi}$.

This property suggests that we define yet another object, this one with \emph{sixteen} components:
\begin{equation}
\Upsilon =  \left( \begin{array}{c} \Phi \\ \hat{\Phi} \end{array} \right) \text{.}
\end{equation}
We also define a conjugate of $\Upsilon$ by
\begin{equation}
\bar{\Upsilon} = \Upsilon^T (i\tau^2)(i\mu^2) \text{.}
\end{equation}
This has introduced yet another $\rSU2$ space into our formulation; we use $\varsigma^i$ Pauli matrices to act in this one.  For example,
\begin{equation}
\varsigma^1 \Upsilon = \left( \begin{array}{c} \hat{\Phi} \\ \Phi \end{array} \right) \text{.}
\end{equation}

The advantage of introducing all this notation is that it allows us to write ${\cal L}_{{\rm eff}}$ in a manifestly ${\rm Sp}(4)$-invariant form.  Some algebra verifies that, up to integration by parts,
\begin{equation}
{\cal L}_{{\rm eff}} = - \frac{1}{2} \bar{\Upsilon} \varsigma^2 \Dslash \Upsilon \text{.}
\end{equation}
This form is invariant under the global rotation
\begin{equation}
\Upsilon \to M \Upsilon \text{,}
\end{equation}
where $M$ should be thought of as a $4 \times 4$ matrix acting in the tensor product of the $\nu^i$ and $\varsigma^i$ Pauli matrix spaces.  (So, for example, $M$ commutes with $\mu^i$ and $\tau^i$ Pauli matrices, but not, in general, $\nu^i$ and $\varsigma^i$.)  The condition that this transformation leaves ${\cal L}_{{\rm eff}}$ invariant is 
\begin{equation}
M^T  \varsigma^2 M = \varsigma^2 \text{.}
\end{equation}
This is precisely the statement that $M$ is an ${\rm Sp}(4)$ matrix, so we have shown that ${\cal L}_{{\rm eff}}$ is invariant under global ${\rm Sp}(4)$ rotations.

One point remains to be addressed.  Because our representation of ${\cal L}_{{\rm eff}}$ is redundant to a certain extent, we have to check that the ${\rm Sp}(4)$ is a real symmetry and not just an artifact of the formalism.  This can be done by rewriting the ${\rm Sp}(4)$ Noether current in terms of the original fermion operators, and checking that it is non-vanishing and that its components are linearly independent.  The Noether current is
\begin{equation}
J^A_\mu =  - \frac{1}{2} \bar{\Upsilon} \gamma_\mu \varsigma^2 T^A \Upsilon \text{,}
\end{equation}
where $T^A$ is one of the ten $4 \times 4$ matrix generators of ${\rm Sp}(4)$. Some straightforward algebra -- or, alternatively, a group-theoretic analysis -- shows that all components of $J^A_\mu$ are nonzero and linearly independent, and therefore the ${\rm Sp}(4)$ is a genuine global symmetry.

\subsection{Microscopic symmetries}

I now enumerate the action of the microscopic symmetries on the continuum fields (in real space).  The results quoted here can be derived starting from the definition of $\varphi_a$ in Eqs.~(\ref{eqn:varphi1}-\ref{eqn:varphi2}).  We need to consider the space group symmetries of the honeycomb lattice, spin-rotation symmetry and time-reversal.  Pseudospin symmetry (including the particle-hole transformation) does not act on the spinon sector, so we do not discuss it here.

First, we discuss the space group symmetries and time-reversal.  (Spin rotations will be dealt with separately.)  For each symmetry, we give its definition in terms of the microscopic spinon fields, then quote its action on the continuum fields $\Upsilon$ and $\bar{\Upsilon}$.  We include enough space group operations to generate the entire space group.

\emph{Translations}.  We consider translations by the basis vectors $\ba_1$ and $\ba_2$.  In general, $T_{\bR} : \psi_{\bR' i} \to \psi_{\bR + \bR', i}$.  In the continuum we have
\begin{eqnarray}
T_{\ba_1}: \Upsilon &\to& \exp \Big( \frac{4 \pi i}{3} \nu^3 \varsigma^3 \Big) \Upsilon \\
T_{\ba_1}: \bar{\Upsilon} &\to& \bar{\Upsilon} \exp \Big( \frac{4\pi i}{3} \nu^3 \varsigma^3 \Big) \\
T_{\ba_2}: \Upsilon &\to& \exp \Big( \frac{2 \pi i}{3} \nu^3 \varsigma^3 \Big) \Upsilon \\
T_{\ba_2}: \bar{\Upsilon} &\to& \bar{\Upsilon} \exp \Big( \frac{2 \pi i}{3} \nu^3 \varsigma^3 \Big) \text{.}
\end{eqnarray}

\emph{Horizontal reflection}.  This operation is defined by ${\cal R}_x : \psi_{\bR i} \to \psi_{\bR' i}$, where $\bR' = (-R_x, R_y)$.  We have
\begin{eqnarray}
{\cal R}_x : \Upsilon(\br) &\to& - \varsigma^3 \nu^2 \tau^2 \Upsilon(\br') \\
{\cal R}_x : \bar{\Upsilon}(\br) &\to& \bar{\Upsilon}(\br') \big[ - \varsigma^3 \nu^2 \tau^2 \big] \text{.}
\end{eqnarray}

\emph{Vertical reflection}.  Here, we have ${\cal R}_y : \psi_{\bR 1} \to \mu^3 \psi_{\bR' 2}$ and
${\cal R}_y : \psi_{\bR 2} \to \mu^3 \psi_{\bR' 1}$, where $\bR' = (R_x, -R_y)$.  The result in the continuum is
\begin{eqnarray}
{\cal R}_y : \Upsilon(\br) &\to& - \varsigma^3 \nu^3 \tau^1 \mu^3 \Upsilon(\br') \\
{\cal R}_y : \bar{\Upsilon}(\br) &\to& \bar{\Upsilon}\big[ - \varsigma^3 \nu^3 \tau^1 \mu^3 \big] \text{.}
\end{eqnarray}

\emph{Rotation about plaquette center}.  We consider a counterclockwise rotation by $\pi/3$ about the plaquette center located at $(1/2,0)$.  The action on the fermion fields is
\begin{eqnarray}
R_{\pi/3} : \psi_{\bR 1} &\to& \mu^3 \psi_{\bR' + \ba_1 - \ba_2, 2} \\
R_{\pi/3} : \psi_{\bR 2} &\to& \mu^3 \psi_{\bR' 1} \text{,}
\end{eqnarray}
where $\bR'$ is the image of $\bR$ under a rotation by $\pi/3$ about the origin.  In the continuum,
\begin{widetext}
\begin{eqnarray}
R_{\pi/3} : \Upsilon(\br) &\to& \varsigma^3  \exp\Big[ - \frac{i \pi}{2} \big( \frac{1}{2}\nu^1 \varsigma^3 + \frac{\sqrt{3}}{2} \nu^2 \big) \Big]  \exp \big( \frac{ i \pi \tau^3}{6} \big) \mu^3 \Upsilon(\br') \\
R_{\pi/3} : \bar{\Upsilon}(\br) &\to& - \bar{\Upsilon}(\br') \varsigma^3
\exp\Big[ - \frac{i \pi}{2} \big( \frac{1}{2}\nu^1 \varsigma^3 - \frac{\sqrt{3}}{2} \nu^2 \big) \Big]
 \exp \big(- \frac{ i \pi \tau^3}{6} \big) \mu^3
 \text{.}
\end{eqnarray}
\end{widetext}

\emph{Time reversal}.  It is simplest to define time reversal in terms of the original $f$-fermions; we have
${\cal T} : f_{\br \alpha} \to (i \sigma^2)_{\alpha \beta} f_{\br \beta}$.  It is important to remember that time reversal is an anti-unitary operation.  Time reversal operates on the continuum fields as follows:
\begin{eqnarray}
{\cal T} : \Upsilon(\br) &\to& - \tau^2 \nu^3 \varsigma^1 \Upsilon(\br) \\
{\cal T} : \bar{\Upsilon}(\br) &\to& \bar{\Upsilon}(\br) \varsigma^1 \tau^2 \nu^3 \text{.}
\end{eqnarray}

Now we consider spin rotations.  As these make up part of the ${\rm Sp}(4)$ symmetry, we proceed somewhat differently from the symmetries above.  First it is useful to provide some more detail about the ${\rm Sp}(4)$ symmetry itself. 
A general ${\rm Sp}(4)$ matrix can be expressed as $M = e^{i \lambda^A T^A}$, where $A = 1,\dots,10$ labels the ${\rm Sp}(4)$ generators, and the condition
$M^T \varsigma^2 M = \varsigma^2$ implies
\begin{equation}
(T^A)^T \varsigma^2 + \varsigma^2 T^A = 0 \text{.}
\end{equation}
We choose the following basis for the $T^A$:
\begin{equation}
T^A = \{ \frac{1}{2} \nu^2, \frac{1}{2} \varsigma^i , \frac{1}{2} \varsigma^i \nu^1, \frac{1}{2} \varsigma^i \nu^3 \} \text{.}
\end{equation}
The conserved charges $Q^A = \int d^2 \br \, J^A_0(\br)$ generate ${\rm Sp}(4)$ rotations.  Formally this means that
\begin{equation}
\label{eqn:sp4-action}
e^{i \lambda^A Q^A} \Upsilon e^{- i \lambda^A Q^A} = e^{- i \lambda^A T^A} \Upsilon \text{,}
\end{equation}
where the objects in this equation should be interpreted as quantum operators, \emph{not} as Grassmann variables.  Actually the form of the right-hand side of this equation is not completely obvious; depending on the normalization chosen for the Noether current, we could have a different coefficient inside the exponential.  To check that Eq.~(\ref{eqn:sp4-action}) is correct, it is enough to evaluate the left-hand side for some particular choice of $\lambda^A$; this is just a matter of some algebra.

To determine the action of spin rotations on $\Upsilon$, the above discussion shows that we need only identify the spin density among the ${\rm Sp}(4)$ conserved densities $J^A_0$.  The correct identification is
\begin{eqnarray}
S^x(\br) &=& J^A_0 \Big[ T^A = \frac{1}{2} \varsigma^2 \nu^1 \Big] \\
S^y(\br) &=& J^A_0 \Big[ T^A = \frac{1}{2} \varsigma^1 \nu^1 \Big] \\
S^z(\br) &=& J^A_0 \Big[ T^A = - \frac{1}{2} \varsigma^3  \Big]  \text{.}
\end{eqnarray}
So, denoting by $R_s(\boldsymbol{\alpha})$ a spin rotation by an angle $|\boldsymbol{\alpha}|$ about the $\boldsymbol{\alpha} / |\boldsymbol{\alpha}|$ axis, we have by Eq.~(\ref{eqn:sp4-action})
\begin{eqnarray}
R_s(\theta \bx) : \Upsilon &\to& \exp \Big( - \frac{i \theta \varsigma^2 \nu^1}{2} \Big) \Upsilon \\
R_s(\theta \boldsymbol{y}) : \Upsilon &\to& \Big( - \frac{i \theta \varsigma^1 \nu^1}{2} \Big) \Upsilon \\
R_s(\theta \boldsymbol{z}) : \Upsilon &\to& \Big( \frac{i \theta \varsigma^3 }{2} \Big) \Upsilon \text{.}
\end{eqnarray}

\subsection{Unification of competing orders}

Here, we construct the gauge-invariant, Lorentz-singlet fermion bilinears in the $\rSU2$ ASL.  There are six such bilinears; five of them transform as the ${\rm SO}(5)$ vector irreducible representation of ${\rm Sp}(4)$, and the other is an ${\rm Sp}(4)$ singlet.  Ran and Wen\cite{ran06} found that the correlations of the ${\rm SO}(5)$ vector bilinears are enhanced by gauge fluctuations (see Sec.~\ref{sec:spin-liquid-ft}).  We show here that the ${\rm SO}(5)$ vector is a unification of the N\'{e}el vector, and the two-component order parameter for the VBS state pictured in Fig.~\ref{fig:hcvbs}.  We also define the VBS order parameter microscopically in terms of spin operators.

A general gauge-invariant, Lorentz-singlet fermion bilinear has the form
\begin{equation}
B_W  = \bar{\Upsilon} W \Upsilon \text{,}
\end{equation}
where $W$ is a $4 \times 4$ matrix acting in the tensor product space of the $\varsigma^i$ and $\nu^i$ Pauli matrices.  We require $B_W$ to be Hermitian, which puts the following condition on $W$:
\begin{equation}
W = \varsigma^2 W^\dagger \varsigma^2 \text{.}
\end{equation}
There are 16 linearly independent matrices satisfying this condition.  We wish to decompose the resulting 16 bilinears into irreducible representations of ${\rm Sp}(4)$.  To do this, note that under an ${\rm Sp}(4)$ rotation $\Upsilon \to M \Upsilon$, $W \to M^T W M$.  This transformation preserves the symmetry of $W$, so we can consider symmetric and antisymmetric $W$ matrices separately.  Furthermore, $\tr (\varsigma^2 W)$ is invariant under ${\rm Sp}(4)$ rotations, so it must be constant within each irreducible representation.  We are therefore led to the following bases of $W$ matrices for the irreducible representations
\begin{eqnarray}
{\cal W}^5_{AS} &=& \{ \varsigma^2 \nu^1, \varsigma^2 \nu^3, \nu^2, i \varsigma^1 \nu^2, i \varsigma^3 \nu^2 \} \\
{\cal W}^1_{AS} &=& \{ \varsigma^2 \} \\
{\cal W}_{S} &=& \{ \varsigma^2 \nu^2, 1, \nu^1, \nu^3, i \varsigma^1, i \varsigma^1 \nu^1, i \varsigma^1 \nu^3 , \nonumber \\
&& i \varsigma^3, i\varsigma^3 \nu^1, i\varsigma^3 \nu^3 \} \text{.}
\end{eqnarray}
The first two of these consist of antisymmetric $W$ matrices; ${\cal W}^5_{AS}$ is the ${\rm SO}(5)$ vector representation, and ${\cal W}^1_{AS}$ is an ${\rm Sp}(4)$ singlet.  It can be seen that all bilinears $B_W$ with the symmetric $W$ matrices of ${\cal W}_S$ vanish identically; this is a consequence of Fermi statistics and the redundancy of our representation for the fermion fields.  This can be seen by showing, for example, that $i \bar{\Upsilon} \varsigma^1 \Upsilon = 0$; all the other bilinears obtained from ${\cal W}_S$ must then vanish by ${\rm Sp}(4)$ symmetry.

Therefore we are left with only six bilinears.  We organize these by defining
\begin{eqnarray}
N^{\cal A} &=& \Big( \bar{\Upsilon} \varsigma^2 (- \varsigma^1 \nu^2) \Upsilon,
\bar{\Upsilon} \varsigma^2 (\varsigma^2 \nu^2) \Upsilon, \bar{\Upsilon} \varsigma^2 (\nu^3) \Upsilon ,\nonumber \\
&& \bar{\Upsilon} \varsigma^2 (\nu^1) \Upsilon, \bar{\Upsilon} \varsigma^2 (\varsigma^3 \nu^2) \Upsilon \Big ) \text{,}
\end{eqnarray}
where ${\cal A} = 1,\dots,5$, and
\begin{equation}
\Omega = \bar{\Upsilon} \varsigma^2 \Upsilon \text{.}
\end{equation}
$N^{\cal A}$ is the ${\rm SO}(5)$ vector and $\Omega$ is the ${\rm Sp}(4)$ singlet.

Next, we define
\begin{equation}
\boldsymbol{n} = (N^1, N^2, N^3) \text{,}
\end{equation}
and
\begin{equation}
\boldsymbol{v} = (N^4, N^5) \text{.}
\end{equation}
It is a straightforward exercise to use the results of the previous section to show that $\boldsymbol{n}$ transforms like the N\'{e}el vector under all microscopic symmetries, and $\boldsymbol{v}$ transforms like the two-component order parameter for the VBS state pictured in Fig.~\ref{fig:hcvbs}.  
 
We now define the VBS order parameter microscopically in terms of spin operators.  It is analogous to the order parameter used by Read and Sachdev to characterize VBS states on the square lattice.\cite{read90}  We define the order parameter as
\begin{eqnarray}
\psi_{{\rm VBS}} &=& \sum_{\bR} \exp \Big( - \frac{4 \pi i}{3} \bR \cdot \ba_1 \Big) \Big[
\bS_{\bR 1} \cdot \bS_{\bR 2}  \nonumber \\ &+&
e^{- 4\pi i / 3} \bS_{\bR 1} \cdot \bS_{\bR + \ba_1 - \ba_2, 2} \nonumber \\ &+&
e^{- 2\pi i / 3} \bS_{\bR 1} \cdot \bS_{\bR - \ba_2, 2} \Big] \text{.}
\end{eqnarray}
The relation between this object and $\boldsymbol{v}$ is
\begin{equation}
\int d^2 \bx \big(v_1(\bx) + i v_2 (\bx) \big) \sim \psi_{{\rm VBS}} \text{.}
\end{equation}

Now consider the pattern of VBS order show in Fig.~\ref{fig:hcvbs}, and suppose that, on the strong bonds, $\langle \bS_{\br} \cdot \bS_{\br'} \rangle = c_1$, while on the weak bonds, $\langle \bS_{\br} \cdot \bS_{\br'} \rangle = c_2$.  Then it can be shown that, in this state,
\begin{equation}
\frac{1}{N_c} \langle \psi_{{\rm VBS}} \rangle = e^{i \theta} \Big[ 3 (c_1 - c_2) \Big] \text{,}
\end{equation}
where $N_c$ is the number of unit cells in the crystal, which should of course be taken to infinity to obtain the thermodynamic limit.  The phase $\theta$ can take on the three values $\theta = 0, \pm 2 \pi/3$, which correspond to the three distinct VBS ground states.

As an aside, it is interesting to note that, if the roles of strong and weak bonds are reversed, the ``bond-ordered'' state of Fig.~\ref{fig:hcvbs} becomes a \emph{plaquette state}, and $\langle \psi_{{\rm VBS}} \rangle$ simply changes sign.  Therefore, the pattern of symmetry breaking is \emph{identical} in the bond-ordered and plaquette states, and it is not at all obvious that these two states are really sharply distinct phases.  On the square lattice, the Read-Sachdev VBS order parameter also contains both the columnar and plaquette VBS states, but these states do not have the property of being related by interchanging strong and weak bonds, and are clearly characterized by different patterns of symmetry breaking.  We note that a study of the honeycomb lattice quantum dimer model\cite{moessner01} found a first-order transition between bond-ordered and plaquette states, which does not exclude the possibility that these states are in fact different limits of the same phase.

\section{Formal equivalence of $\rSU2$ and $\rU1$ mean-field calculations}
\label{app:mft-equiv}
 
Here, we demonstrate the formal equivalence of the mean-field calculations for the $\rSU2$ and $\rU1$ slave rotor formulations of the Hubbard model.  This equivalence holds for the restricted class of mean-field ansatz for the $\rSU2$ formulation given in Eqs.~(\ref{eqn:ansatz-1}-\ref{eqn:ansatz-2}).  We may therefore use the results of Ref.~\onlinecite{sslee05} for the honeycomb Hubbard model in our discussion of the mean-field phase diagram in Sec.~\ref{sec:mft-phasediag}.

Below, we shall consider  saddle points with $\tilde{\ba}_\tau = 0$.  In that case, the ground state energy in mean-field theory is given by
\begin{eqnarray}
E_0^{{\rm MFT}} &=& \sum_{( \br, \br' ) } |t_{\br \br'}| |\chi^F_{\br \br'}| |\chi^Z_{\br \br'}| \tr ( U^{\br' \br}_F U^{\br \br'}_Z ) \nonumber  \\ &-& \sum_{\br} \tilde{\lambda}_{\br} + E_0^F + E_0^Z \label{eqn:mft-energy-1}\text{,}
\end{eqnarray}
where $E_0^F$ and $E_0^Z$ are the ground state energies of the fermions and bosons, respectively.  The fermion energy can easily be obtained by taking the expectation value of the fermion mean-field Hamiltonian:
\begin{eqnarray}
E_0^F &=& \sum_{( \br, \br' )} t_{\br \br'} |\chi^F_{\br \br'}| \tr \big( F_{\br'} U^{\br' \br}_F F^\dagger_{\br} \big) \\
&=& - \sum_{(\br, \br')} |t_{\br \br'}| |\chi^F_{\br \br'}| |\chi^Z_{\br \br'}| \tr \big( U^{\br' \br}_F U^{\br \br'}_Z \big) \text{.}
\end{eqnarray}
This cancels the first term of Eq.~(\ref{eqn:mft-energy-1}), and
\begin{equation}
\label{eqn:mft-energy-2}
E^{{\rm MFT}}_0 = E_0^Z - \sum_{\br} \tilde{\lambda}_{\br} \text{.}
\end{equation}
 
To demonstrate the  equivalence of the present mean-field calculation and that of Ref.~\onlinecite{sslee05}, it is first useful to plug our ansatz into the saddle point action.  The result for the rotor boson term is
\begin{eqnarray}
\tilde{S}_Z &=& \int d\tau \Big[ \sum_{\br} z^*_{\br i} \big( - \frac{3}{2 U} \partial_\tau^2 + \tilde{\lambda}_{\br} \big) z_{\br i}  \nonumber \\ &-& \sum_{(\br , \br')} | t_{\br \br'} | |\chi^Z_{\br \br'} | \big( e^{i \tilde{a}^Z_{\br \br'} } z^*_{\br i} z_{\br' i} + \text{c.c} \big) \Big] \\
&=& \int d\tau \sum_\gamma z^*_{\gamma i} \Big( - \frac{3}{2 U} \partial_\tau^2 + \epsilon_\gamma \Big) z_{\gamma i}  \label{eqn:su2-zaction-diagonalform} \text{,}
\end{eqnarray}
where in the second equality we have transformed to a basis labeled by $\gamma$ that diagonalizes the quadratic form, with eigenvalues $\epsilon_\gamma$.  Explicitly,
\begin{equation}
\label{eqn:defn-of-eigenfunctions}
z_{\br i} = \sum_\gamma f_\gamma (\br) z_{\gamma i} \text{.}
\end{equation}
The fermion part of the saddle-point action is
\begin{eqnarray}
\tilde{S}_F &=& \int d\tau \Big[ \sum_{\br} \bar{f}_{\br \alpha} \partial_\tau f_{\br \alpha}
  \nonumber \\ &-& \sum_{(\br, \br')} t_{\br \br'} |\chi^F_{\br \br'}| \big( e^{-i \tilde{a}^F_{\br \br'} } 
\bar{f}_{\br \alpha} f_{\br' \alpha} + \text{H.c.} \big)  \Big] \text{.}
\end{eqnarray}
Using Eq.~(\ref{eqn:mft-energy-2}), the mean-field ground state energy is
\begin{equation}
\label{eqn:su2-gsenergy}
E^{{\rm MFT}}_0 = 2 \sqrt{\frac{2 U}{3}} \sum_{\gamma} \sqrt{\epsilon_\gamma}\,\, - \,\, \sum_{\br} 
\tilde{\lambda}_{\br} \text{.}
\end{equation}
The saddle-point equations become
\begin{eqnarray}
1/2 &=& \langle | z_{\br 1} |^2 \rangle = \langle |z_{\br 2}|^2 \rangle  \\
0 &=& \langle \psi^\dagger_{\br} \bsigma \psi^{\vphantom\dagger}_{\br} \rangle \\
|\chi^Z_{\br \br'}| e^{i \tilde{a}^Z_{\br \br'} } &=& \operatorname{sign}(t_{\br \br'}) \langle f^\dagger_{\br \alpha} f_{\br' \alpha} \rangle \\
|\chi^F_{\br \br'}| e^{-i \tilde{a}^F_{\br \br'} } &=& 2 \langle z^*_{\br 1} z_{\br' 1} \rangle = 
2 \langle z^*_{\br 2} z_{\br' 2} \rangle \text{,}
\end{eqnarray}
and the remaining equation, $\operatorname{Im} \langle \operatorname{tr} ( \partial_\tau Z^\dagger_{\br} \bsigma Z_{\br} ) \rangle = 0$, is automatically satisfied by saddle points of the form considered.

Now we will briefly introduce the $\rU1$ formulation, and then show that it reproduces the same saddle-point equations and ground state energy.  Following Ref.~\onlinecite{sslee05}, the partition function of the Hubbard model can be written
\begin{equation}
Z = \int {\cal D} f {\cal D} \bar{f} {\cal D} X {\cal D}X^* {\cal D} h {\cal D} \lambda \exp \big( - S_{{\rm U}(1)} \big) \text{,}
\end{equation}
where $S_{ \rU1} = \int d\tau L_{ \rU1 } (\tau)$ and
\begin{eqnarray}
L_{ \rU1 } &=& \frac{1}{2 U'} \sum_{\br}  \big| ( i \partial_\tau + h_{\br} ) X_{\br} \big|^2
+ \sum_{\br} \bar{f}_{\br \alpha} (\partial_\tau + i h_{\br} ) f_{\br \alpha} \nonumber \\
&-& i \sum_{\br} h_{\br} + i \sum_{\br} \lambda_{\br} ( X^*_{\br} X_{\br} - 1) \nonumber \\
&-& \sum_{(\br , \br') } \big( t_{\br \br'} \bar{f}_{\br \alpha} f_{\br' \alpha} X^*_{\br} X_{\br'} + \text{H.c.} \big) \text{.}
\end{eqnarray}
Here it is important to note that we have taken the coefficient of the on-site repulsive interaction to be $U'$ rather than $U$.  It will turn out that the $\rU1$ and $\rSU2$ mean-field calculations are equivalent provided that $U' = (4/3) U$.

The hopping term is decoupled using a pair of complex fields $\eta_{\br \br'}$ and $\eta_{\br' \br}$, resulting in the following contributions to the action:
\begin{eqnarray}
S^{ \rU1}_\eta &=& \int d\tau \sum_{(\br, \br') } |t_{\br \br'}| \big( | \eta_{\br \br'} |^2 + | \eta_{\br' \br} |^2 \big) \\
S^{\rU1}_{tX} &=& - \sum_{(\br, \br')} |t_{\br \br'}| \big ( X^*_{\br} \eta_{\br \br'} X_{\br'} \nonumber \\
&+& X^*_{\br'} \eta_{\br' \br} X_{\br} \big) \\
S^{\rU1}_{tf} &=& - \int d\tau \sum_{(\br, \br') } t_{\br \br'} \big( \bar{f}_{\br \alpha} \eta^*_{\br \br'} f_{\br' \alpha} \nonumber \\ &+& \bar{f}_{\br' \alpha} \eta^*_{\br' \br} f_{\br \alpha} \big ) \text{.}
\end{eqnarray}
We can now look for saddle points by varying the action with respect to $\eta_{\br \br}$, $\eta_{\br' \br}$, $\lambda_{\br}$ and $h_{\br}$.  The resulting saddle-point equations have the form
\begin{eqnarray}
\eta^*_{\br \br'} &=& \langle X^*_{\br} X_{\br'} \rangle \\
\eta_{\br \br'} &=& \operatorname{sign}(t_{\br \br'}) \langle \bar{f}_{\br \alpha} f_{\br' \alpha} \rangle \\
1 &=& \langle X^*_{\br} X_{\br} \rangle \\
0 &=& \frac{1}{U'} h_{\br} + i \langle \bar{f}_{\br \alpha} f_{\br \alpha} \rangle
+ \frac{1}{U} \big\langle (\partial_\tau X^*_{\br} ) X_{\br} \big\rangle \text{,}
\end{eqnarray}
where we have not explicitly written the equations involving $\eta_{\br' \br}$, which are similar in form to those involving $\eta_{\br \br'}$.  We now write
\begin{eqnarray}
\eta_{\br \br'} &=& |\chi_{\br \br'} | e^{w_{\br \br'}} e^{i ( a^+_{\br \br'} + a_{\br \br'} )} \\
\eta_{\br' \br} &=& |\chi_{\br \br'} | e^{- w_{\br \br'}} e^{i ( a^+_{\br \br'} - a_{\br \br'} )} \text{,}
\end{eqnarray}
and make the analytic continuation
\begin{eqnarray}
w_{\br \br'} &\to& i \tilde{w}_{\br \br'}  \\
a^+_{\br \br'} &\to& i \tilde{a}^+_{\br \br'} \\
\lambda_{\br} &\to& -i \tilde{\lambda}_{\br}  \\
h_{\br} &\to& i \tilde{h}_{\br} + \delta h_{\br} \text{.}
\end{eqnarray}

We look for saddle points where $\tilde{h}_{\br} = 0$.  Then, defining 
\begin{eqnarray}
|\chi^X_{\br \br'}| &\equiv& |\chi_{\br \br'} | e^{- \tilde{a}^+_{\br \br'} }  \\
|\chi^f_{\br \br'}| &\equiv& |\chi_{\br \br'} | e^{ \tilde{a}^+_{\br \br'} }\\
\tilde{a}^X_{\br \br'} &\equiv& \tilde{w}_{\br \br'} + a_{\br \br'} \\
\tilde{a}^f_{\br \br'} &\equiv& \tilde{w}_{\br \br'} - a_{\br \br'} \text{,}
\end{eqnarray}
the saddle-point equations are
\begin{eqnarray}
|\chi^X_{\br \br'} | e^{i \tilde{a}^X_{\br \br'}}  &=& \operatorname{sign}(t_{\br \br'}) \langle \bar{f}_{\br \alpha} f_{\br' \alpha} \rangle  \label{eqn:u1-f-spe} \\
|\chi^f_{\br \br'} | e^{-i \tilde{a}^f_{\br \br'} }&=& \langle X^*_{\br} X_{\br'} \rangle \label{eqn:u1-X-spe} \\
1 &=& \langle X^*_{\br} X_{\br} \rangle  \label{eqn:u1-sigmamodel-spe} \\
1 &=& \langle \bar{f}_{\br \alpha} f_{\br \alpha} \rangle  \label{eqn:u1-half-filling-spe} \\
0 &=& \operatorname{Im} \langle (\partial_\tau X^*_{\br} ) X_{\br} \rangle \label{eqn:u1-impart-spe} \text{.}
\end{eqnarray}
As in the $\rSU2$ case, Eq.~(\ref{eqn:u1-impart-spe}) is trivially satisfied.
The expectation values are to be calculated using the saddle-point action $\tilde{S}_{ \rU1} = \int d\tau \tilde{L}_{\rU1}(\tau)$, where
\begin{eqnarray}
\tilde{L}_{ \rU1} &=& 2 \sum_{(\br, \br')} |t_{\br \br'}| |\chi^f_{\br \br'}| |\chi^X_{\br \br'}| \cos(\tilde{a}^X_{\br \br'} - \tilde{a}^f_{\br \br'} ) - \sum_{\br} \tilde{\lambda}_{\br} \nonumber \\
&+& \sum_{\br} \bar{f}_{\br \alpha} \partial_\tau f_{\br \alpha} + \sum_{\br} X^*_{\br} \Big( - \frac{1}{2 U'} \partial_{\tau}^2 + \tilde{\lambda}_{\br} \Big) X_{\br} \nonumber \\
&-& \sum_{(\br, \br' )} t_{\br \br'} |\chi^f_{\br \br'}| \big( e^{-i \tilde{a}^f_{\br \br'} }\bar{f}_{\br \alpha} f_{\br' \alpha} + \text{H.c.} \big) \nonumber \\
&-& \sum_{(\br, \br')} | t_{\br \br'} | |\chi^X_{\br \br'}| \big( X^*_{\br} e^{i \tilde{a}^X_{\br \br'} } X_{\br'} + \text{c.c.} \big)
\label{eqn:u1-spoint-action} \text{.}
\end{eqnarray}

If we identify $|\chi_{\br \br'}^f| = |\chi_{\br \br'}^F|$ and $\tilde{a}^F_{\br \br'} = \tilde{a}^f_{\br \br'}$, the fermion part of Eq.~(\ref{eqn:u1-spoint-action}) is identical to $\tilde{S}_F$ of the $\rSU2$ formulation.  All expectation values of fermion operators are then equal in the two formulations, and by Eq.~(\ref{eqn:u1-f-spe}) we should also identify $|\chi_{\br \br'}^X| = |\chi_{\br \br'}^Z|$ and $\tilde{a}^X_{\br \br'} = \tilde{a}^Z_{\br \br'}$.  So far we have shown that the saddle-point Eqs.~(\ref{eqn:u1-f-spe}) and~(\ref{eqn:u1-half-filling-spe}) are identical to their counterparts in the $\rSU2$ formulation.

The part of Eq.~(\ref{eqn:u1-spoint-action}) involving $X_{\br}$ can be written
\begin{eqnarray}
\tilde{L}^{\rU1}_X &=& \sum_{\gamma} X^*_{\gamma} \Big( - \frac{1}{2U'} \partial_{\tau}^2 + \epsilon_\gamma \Big) X_{\gamma} \text{,}
\end{eqnarray}
where
\begin{equation}
X_{\br} = \sum_{\gamma} f_{\gamma} (\br) X_{\gamma}
\end{equation}
and $f_\gamma (\br)$ and $\epsilon_\gamma$ are the \emph{same} eigenfunctions and eigenvalues as in Eqs.~(\ref{eqn:su2-zaction-diagonalform}) and~(\ref{eqn:defn-of-eigenfunctions}). Proceeding as above, the ground state energy is
\begin{equation}
E^{{\rm MFT}}_0 = \sqrt{2 U'} \sum_{\gamma} \sqrt{\epsilon_\gamma} - \sum_{\br} \tilde{\lambda}_{\br} \text{.}
\end{equation}
This can be made equal to the ground state energy calculated in the $\rSU2$ formulation, Eq.~(\ref{eqn:su2-gsenergy}), by choosing
\begin{equation}
U' = \frac{4}{3} U \text{.}
\end{equation}

We need to check that this choice for $U'$ will also make the remaining saddle-point equations, Eqs.~(\ref{eqn:u1-X-spe}) and~(\ref{eqn:u1-sigmamodel-spe}),  identical to their $\rSU2$ counterparts.  To see this, we evaluate the equal-time correlator (in the $\rSU2$ mean-field theory)
\begin{eqnarray}
\langle z^*_{\br 1} z_{\br' 1} \rangle &=& \sum_\gamma \sum_{\gamma'} f^*_{\gamma}(\br) f_{\gamma'}(\br')
\langle z^*_{\gamma 1} z_{\gamma' 1} \rangle \\
&=& \int_{-\infty}^{\infty}  \frac{d \omega_n}{ 2\pi} \sum_{\gamma} \frac{f^*_\gamma (\br) f_\gamma (\br')}
{ (3/ 2 U) \omega_n^2 + \epsilon_\gamma} \text{.}
\end{eqnarray}
Similarly, in the $\rU1$ formulation, putting $U' = (4/3)U$, we have
\begin{eqnarray}
\langle X^*_{\br} X_{\br'} \rangle &=& \int_{-\infty}^{\infty} \frac{d \omega_n}{2\pi} \sum_{\gamma}
\frac{f^*_\gamma (\br) f_\gamma (\br')}
{ (3/ 8 U) \omega_n^2 + \epsilon_\gamma} \\
&=& 2  \int_{-\infty}^{\infty} \frac{d \omega_n}{2\pi} \sum_{\gamma}
\frac{f^*_\gamma (\br) f_\gamma (\br')}
{ (3/ 2 U) \omega_n^2 + \epsilon_\gamma}  \text{,}
\end{eqnarray}
where in the last line we have changed variables by $\omega_n \to 2 \omega_n$.  Therefore we have shown that
\begin{equation}
\langle X^*_{\br} X_{\br'} \rangle = 2 \langle z^*_{\br 1} z_{\br' 1} \rangle = 2 \langle z^*_{\br 2} z_{\br' 2} \rangle \text{,}
\end{equation}
which implies that the remaining saddle-point equations are identical in the two formulations.

\bibliography{su2}

\end{document}